\documentclass[nosumlimits]{amsart}

\usepackage{amsmath}
\usepackage{amssymb}

\newtheorem{thm}{Theorem}[section]
\newtheorem{defin}[thm]{Definition}

\newtheorem{cor}[thm]{Corollary}

\newtheorem{pro}[thm]{Proposition}
\numberwithin{equation}{section}

\newcommand{\E}{\mathcal{E}}  
\newcommand{\lip}{\textit{l}}  
\newcommand{\D}{\mathcal{D}}  
\newcommand{\borel}{\mathcal{B}}  

\newcommand{\A}{\mathcal{A}}
\newcommand{\T}{\mathcal{T}}

 \newcommand{\into}{\rightarrow}
 \newcommand{\impl}{\Longrightarrow}
 
 \newcommand{\abs}[1]{\left\vert#1\right\vert}
 \newcommand{\set}[1]{\left\{#1\right\}}
 
 \newcommand{\dotp}[2]{\left<#1,#2\right>}

 \newcommand{\norm}[1]{\left\Vert#1\right\Vert}
 
 \newcommand{\qtext}[1]{\quad\text{#1}\quad}
 \newcommand{\fa}{\qtext{for all}}
 \newcommand{\bb}{\begin{equation*}}
 \newcommand{\ee}{\end{equation*}}
 \newcommand{\bp}{\begin{proof}}
 \newcommand{\ep}{\end{proof}}

\begin{document}

\title[Fields generated by moving point mass]{Fields generated by\\ a moving
 relativistic point mass and\\
mathematical correction to Feynman's law}

\author{ Victor M. Bogdan }

\address{Department of Mathematics, McMahon Hall 207, CUA, Washington DC 20064, USA}



\email{bogdan@cua.edu}

%
%
\dedicatory{This paper is dedicated to my teachers: Sir Isaac Newton,
Gottfried W. Leibnitz, Rudolf O. S. Lipschitz, James C. Maxwell,
Hendrik A. Lorentz, Albert Einstein, Stefan Banach, Stanis{\l}aw Mazur,
Richard P. Feynman, and Stan M. Ulam, who directly or indirectly inspired and influenced
this research.}

\begin{abstract}
%
%
 Feynman using a formula, known as Feynman's law  for a moving point charge,
explained the phenomenon of  synchrotron radiation
and derived formulas for phenomena,
concerning electromagnetic radiation at large distance from the source, such as
reflection, refraction, interference,
diffraction, and scattering. These facts show the importance of the formula.

 The formula is supposed to represent the intensities of the electric and magnetic
fields in free space, that is satisfying homogeneous system of Maxwell equations.

 Feynman's law contains a mathematical inconsistency. It involves implicitly
a field representing the retarded time and an ordinary derivative
in place where a partial derivative should be.

 In this note the author shows how to prove, using Banach's contraction mapping theorem,
the existence and uniqueness of the retarded time field for
any relativistically admissible trajectory of a point mass.

  The Lorentzian frame, the trajectory, and the retarded time field uniquely
determine a system of fundamental fields.

  By means of these fields one can represent and establish relations between
the system of wave-gauge equations and Maxwell equations, and to prove that
electromagnetic field represented by the amended Feynman's formula
satisfies the homogenous system of Maxwell equations and the system of wave-gauge equations.

As applications the author proves the existence and uniqueness of the
solution to the n-body problem in the resulting joint gravitational and electromagnetic fields
as also in any other relativistic force field
representing a regular nonanticipating operator of the system of trajectories.
\end{abstract}


\maketitle

\section{Analysis of the original Feynman's formula\\
 for a moving point charge}
\bigskip

Importance of Maxwell's laws for logical analysis of
electromagnetic phenomena stems from Einstein's work on
special theory of relativity \cite{einstein2a}.
An asymmetry in the interpretation of an electromagnetic phenomena
based on Lorentz force \cite{lorentz1}, acting on a charged particle, lead
Einstein to the discovery of this theory.

Einstein proved that laws of physics have to be invariant
under transformations discovered earlier by Lorentz \cite{lorentz1}.
In the process he established that Maxwell's equations
are invariant under such transformations \cite{einstein2a}.

Feynman derived heuristically a formula for the electromagnetic field generated by
a moving point charge.
In this paper we will prove that after replacing the ordinary derivative
with respect to time by the partial derivative and treating all quantities
appearing in Feynman's formula as fields, the corrected
formula yields electromagnetic field $(E,B)$ satisfying Maxwell equations.
In the process we shall also prove that  Li\'{e}nard-Wiechert formulas
provide an electromagnetic potential for such a field.

To see that one has to be careful when treating derivatives
\begin{equation*}
    \frac{d}{dt}\qtext{and}\frac{\partial}{\partial t}
\end{equation*}
of composed function consider the following example. Let $u=u(x,y,z,t)$
be a differentiable function in $R^4$ representing some physical
quantity as a function of position and time.
Let $(x(t),y(t),z(t))$ denote a path of a particle as a function of time.
Then the derivative with respect to time along the path is
\begin{equation*}
 \frac{d}{dt}u=u_x\,\dot{x}+u_y\,\dot{y}+u_z\,\dot{z}+u_t.
\end{equation*}
Obviously it is not the same as partial derivative on the path
\begin{equation*}
 \frac{\partial}{\partial t}u=u_t.
\end{equation*}
\bigskip

Any moving point mass $m_0$ obeying Einstein's laws
of special theory of relativity generates a pair of fields $(E,B)$
satisfying Maxwell's equations just from geometrical considerations
of the Lorentzian frame. No other
physical properties are required.
Thus the obtained field $E$ may as well represent
for instance the illusive gravity field that Einstein was looking for
\cite{einstein4}.

To formulate briefly the main result of the paper assume that
in a given Lorentzian frame we introduce fundamental fields that are uniquely
determined by the trajectory of a moving point mass.
By means of these fields and their partial derivatives we can
represent in explicit form fields given by amended
Feynman's and Li\'{e}nard-Wiechert's formulas and
prove the relations between such fields and Maxwell's and wave equations.

The main result can be  stated as follows:
\begin{thm}[Bogdan-Feynman Theorem]
Assume that in a Lorentzian frame is given an   admissible
trajectory $t\mapsto r_2(t)$ from $R$ into $R^3$
representing a path of a point mass $m_0.$
Let $G$ denote  the set of points that do
not lie on the path.
Define Newton-Feynman field by the formula
\begin{equation*}
    E(r_1,t)=u^2e+\frac{1}{c}\,u^{-1}\frac{\partial }{\partial t}(u^2e)+
        \frac{1}{c^2}\frac{\partial^2 }{\partial t^2}(e)\fa r_1\in R^3,\ t\in R
\end{equation*}
and the associated field by
\begin{equation*}
    B(r_1,t)=\frac{1}{c}\,e\times E\fa r_1\in R^3,\ t\in R.
\end{equation*}

Every pair of such
fields $(E,B)$ satisfies the homogenous system of Maxwell equations
and the homogenous system of wave equations on the set $G.$
As such these fields propagate
through space with velocity $c$ of light.

The 4-vector obtained by Li\'{e}nard-Wiechert formulas provides
a pair of scalar-vector potentials for the pair $(E,B).$

\end{thm}
\bigskip
For the notion of an admissible trajectory
see (\ref{admissible trajectory}) and
for definitions of the involved fields $u$ and $e$ see
(\ref{Fundamental fields}).
\bigskip

We shall start with analysis of Feynman's original formulas:
\bigskip

{\em
The intensity of the electric field $E$
and of the magnetic field $B$ at any time $t$ and any point $r_1\in R^3,$ not
lying on the trajectory of  a moving charge $q,$
are given by
\begin{equation}\label{Feynman formulas}
 E=\frac{q}{4\pi\epsilon_0}\left[\frac{e}{|r|^2}
+\frac{|r|}{c}\frac{d}{dt}\left(\frac{e}{|r|^2}\right)
+\frac{1}{c^2}\frac{d^2}{dt^2}\ e\right],\quad
B=\frac{1}{c}\ e\times E,
\end{equation}
where $r=r_1-r_2(t')$ is a vector, starting on the trajectory $t\mapsto r_2(t)$ of a
moving point charge at the retarded time $t'=t-|r|/c$ and ending at the
point $r_1$ at time $t$ where the field is to be evaluated. Here
 $e$ denotes the unit vector corresponding to the vector $r,$ and
 $c$ the speed of light, and $\epsilon_0$ the electrostatic constant.
}

For reference see Feynman-Leighton-Sands
\cite{feyn2}, vol. 1, chapter 28, formulas (28.3) and (28.4).
\bigskip

Let us pose for a moment to analyze the formulas
\ref{Feynman formulas} for their mathematical
content.
On the left side of the formula we have the quantity $E$ representing intensity
of the electric field at point $r_1$ at time $t.$ Thus $E=E(r_1,t)$ represents
a function of the point $(r_1,t).$ So on the right side we should also
find an expression representing a function of these variables.
The vector $r=r_1-r_2(t')$ and  therefore the vector $e=r/|r|$
involves a retarded time that must satisfy the equation
\begin{equation*}
    t'=t-|r|/c
\end{equation*}
equivalently
\begin{equation*}
    t'=t-|r_1-r_2(t')|/c
\end{equation*}
Since the trajectory $t\mapsto r_2(t)$ is fixed the above equation implicitly
defines the retarded time as a function of $(r_1,t).$ If one can solve the
above equation explicitly we will get $t'=t'(r_1,t)$ in the form of a function.
Once this function is known and we can prove that it has continuous partial
derivatives up to order 2 and the trajectory $r_2(t)$ itself has
continuous derivative up to order 2, then the composed functions
\begin{equation*}
 r(r_1,t)=r_1-r_2(t'(r_1,t))\qtext{and}e(r_1,t)=\frac{1}{|r(r_1,t)|}r(r_1,t)
\end{equation*}will be well defined on the set
$G=\set{(r_1,t):\ |r(r_1,t)|>0}$ and the functions
will be of class $C^2.$
\bigskip

Our immediate goal will be to find an explicit formula for the retarded time
function and to prove that the functions appearing in Feynman's formula
indeed have the required properties.

To simplify the notation we select units of measure so that the speed
of light is $c=1$ and the electrostatic
constant satisfies the condition $4\pi\epsilon_0=1.$
We shall also assume that all the intensities of the involved
fields are per unit of charge, that is we assume that $q=1.$
Let us introduce also function $u=1/|r|.$

After these modifications and clarifications the formula for $E$ can be reduced to
a formula

\begin{defin}[Newton-Feynman formula]
\begin{equation}\label{Newton-Feynman formula}
 E(r_1,t)={u^2}{e}
+u^{-1}\frac{\partial }{\partial t}(u^2e)
+\frac{\partial ^2}{\partial t^2}\, e\fa (r_1,t)\in G.
\end{equation}
\end{defin}

Notice that, when the point mass representing the charge, is in its rest frame,
that is $\dot{r}_2(t)=0$ for all $t,$ then the above formula reduces to
a single term $$u^2e=\frac{1}{|r_1-r_2|^3}(r_1-r_2)$$ differing from Newton's
gravitation formula just by a constant of proportionality. Thus it is proper to call
the expression (\ref{Newton-Feynman formula}) the {\bf Newton-Feynman formula.}

\section{Considerations concerning trajectories}
\bigskip

In this section we shall refine the notion of the trajectory $t\mapsto r_2(t)$
so that we could
prove that the functions involved in amended Feynman's formulas are of class $C^3$
that is the functions and their partial derivatives up to order 3 are continuous
on the set $G$ of points not lying on the trajectory. We will need this class of
regularity in order to be able to prove that the  fields generated by amended
Feynman's
formula coincide with fields generated by means of  Li\'{e}nard-Wiechert potentials.

A trajectory of a path of a point mass can be parameterized in several different
ways. It is important to understand which of these parameterizations depend on the
Lorentzian frame, which are invariant under Lorentzian transformations and thus
belong to Einstein's special theory of relativity, and which can be carried over
to general theory of relativity.

Following Einstein \cite{einstein2a} we define a {\bf Lorentzian frame} to consist
of an orthogonal coordinate system in $R^3,$
having right hand orientation of axes, and equipped with a clock.
A physical event in such a frame is described by a point $(r_1,t),$
where  $r_1$ denotes a position in $R^3$ and $t\in R$ the time of the event.
Denote such a frame by $S.$ Assume that $S'$ denotes another
Lorentzian frame whose origin initially coincides with the origin of the frame $S.$
Moreover the frame  $S'$ moves as a rigid body
away from the frame of $S$ at a constant velocity.
The transformation  of coordinates of events from the frame $S$ into the frame $S'$
forms a linear transformation that preserves the quadratic form
\begin{equation*}
    |r'_1|^2-(t')^2=|r_1|^2-t^2.
\end{equation*}

By {\bf geometry of Lorentz space-time} we shall understand the product space
$R^3\times R$ with the transformations of coordinates as described above.
These transformations form a group with composition of transformations as a group operation.
Though one could expand the group by adding affine transformations,
the linear transformations are sufficient for description of dynamics in physical processes
in Einstein's special theory of relativity.
Any affine orthogonal transformation can be reduced
to a linear one just by moving the origin of the coordinate system.

More general groups of transformations related to Lorentz group were studied by
several authors. For generalization of such transformations and further references
see Vogt \cite{vogt}.

Let $\alpha\mapsto y(\alpha)$ be a mapping of an interval
$I$ into $R^4,$ of class $C^3,$
that is having continuous derivatives up to order $3$ on the entire interval $I.$
Assume that the mapping forms a parametric representation of a {\bf path of a point
mass} in space.

Moreover assume that the tangent vector field $y'$ consists of {\bf time-like vectors}
that is
\begin{equation}\label{y' is time like}
    y'_1(\alpha)^2+y'_2(\alpha)^2+y'_3(\alpha)^2-y'_4(\alpha)^2<0\fa \alpha \in R.
\end{equation}
As a derivative of a covariant field  with respect to a free parameter the
tensor $y'_j(\alpha)$ itself forms a covariant field over $I.$
Thus it is invariant under Lorentzian transformations and it can be carried over to
the general theory of relativity as in Dirac \cite{dirac}.
The transition from covariant to contravariant tensors is given by
means of transformation
\begin{equation*}
    y^j=g^{jk}y_k
\end{equation*}
where summation is with respect to index $k=1,2,3,4$
and the matrix $g^{jk}$ for orthogonal axes has elements on the diagonal
equal respectively $1,1,1,-1$ and non-diagonal elements are zero.

The time along the path is given by $t=y_4(\alpha),$ since
from the relation (\ref{y' is time like}) follows that that
$\frac{dt}{d\alpha}=y'_4(\alpha)>0$
for all $\alpha\in I,$ the correspondence $\alpha\mapsto t$ represents a diffeomorphism
of $I$ onto some interval $J$ and is also of class $C^3,$
that is both maps $\alpha\mapsto t=y_4(\alpha)$ and its
inverse $t\mapsto \alpha$ are of class $C^3.$

Thus in every Lorentzian frame we can represent our path in the form
\begin{equation*}
    y=(r_2(t),t)\fa t\in J,
\end{equation*}
where $t\mapsto r_2(t)$ is from some interval $J$ into $R^3$
and represents the position of the mass
as a function of time $t$ in that Lorentzian frame. Clearly this representation
is also of class $C^3$ and forms another equivalent parametric representation of the path
but this representation is frame dependent.

Most important parametrization of a path is with respect the
{\bf proper time}
$s$ of the moving mass $m_0.$ It is unique up to an additive constant and can be found
from the formula
\begin{equation}\label{proper time}
    (ds)^2=(dy_4)^2-\left((dy_1)^2+(dy_2)^2+(dy_3)^2\right)=(dt)^2-|dr_2|^2.
\end{equation}

The condition \ref{y' is time like} can be translated into
\begin{equation*}
    |\dot{r}_2(t)|=\left|\frac{dr_2(t)}{dt}\right|<1=c\fa t\in J,
\end{equation*}
that is velocity along any path of a point mass is less then the speed $c$ of light.
The proper time of a body carries over to general theory of relativity and thus
it is also invariant under Lorentzian transformations.

Maxwell established that waves in electromagnetic field
propagate with velocity of light $c.$
From considerations of Einstein and Rosen \cite{einstein4} follows that even
disturbances in gravity field should propagate with velocity of light.

From results of Bogdan \cite{bogdan61} and
\cite{bogdan64}, Proposition 5.2, follows that if we consider the dynamics of
$n$ bodies interacting with each other by means of fields propagating with
velocity of light, the equations of evolution are non-anticipating differential
equations and their solutions, not only depend on the initial conditions like
in Newtonian mechanics, but also on the initial trajectory of the entire system.

Assuming for instance that in a Lorentzian frame
we are starting with $n$ bodies whose initial trajectories
$t\mapsto y_j(t),$ where $j=1,\ldots,n,$ are known and we intend to observe
the dynamics of evolution of the system for a period of time $t_1,$ and we can
apriori estimate the bound $v$ on velocities, and the bound $A$ on the accelerations
and the initial diameter $\delta$ of the system, then the length of the interval
of significance is at most, according to Proposition 5.2 of \cite{bogdan64},
\begin{equation*}
    a=(\delta+2vt_1)/(c-v).
\end{equation*}
Thus it is sufficient to know the initial trajectories of the system
on the closed interval $[a,0].$

We should think about such trajectories as a postmortem record of the trajectory
of some particular body from the system.
It is clear that such trajectories would correspond to
a time interval $J$ that on the left is closed and on the right open or closed,
finite or infinite. In any case it suffices to restrict ourselves to trajectories
defined on intervals of the form $J=[a,b)$ closed on the left and open on the
right. The left end $a$ of such time interval will be called a {\bf point of significance.}
Any time $t_1$ inside of the interval will be called a {\bf stopping time.}

Thus, if our trajectory $t\mapsto r_2(t)$ is of class $C^2,$ from continuity of
the velocity $w(t)=\dot{r}_2(t)$ and of acceleration $\dot{w}(t)$ on the closed
interval $[a,t_1]$ follows that the following two functions
\begin{equation}\label{q and A boundes}
    \begin{split}
    q(t_1)&=\sup\set{|w(u)|:\ u\le t_1, u\in J}<c,\\
    A(t_1)&=\sup\set{|\dot{w}(u)|:\ u\le t_1, u\in J}<\infty,\\
\end{split}
\end{equation}
are well defined for all stopping times $t_1\in J,$ since the supremum of a continuous
function on a closed bounded interval is attained at some point of that interval.

For the sake of mathematical simplicity we shall consider only trajectories defined on
the entire interval $(-\infty,\infty)=R.$

\begin{defin}[Admissible trajectory]
\label{admissible trajectory}
Assume that we are given a path of a point mass $m_0$ that in some Lorentzian frame
has a representation in the form $y=(r_2(t),t),$ where the
function $r_2(t)$ is from $R$ into $R^3$ and it
has continuous derivatives up to order 3 and that for any stoping time $t_1\in J$
the kinetic energy and the acceleration $\ddot{r}_2(t)$ are bounded on the
interval $(-\infty,t_1\rangle.$
We shall say that such a function $r_2(t)$ represents an {\bf
admissible trajectory.}
\end{defin}

For the sake of logical completeness we should prove that every trajectory with
a point of significance can be extended on the left to an admissible trajectory.
Let us skip this for present and concentrate on admissible trajectories.

\begin{pro}[Kinetic energy bound and velocity bound]
Assume that a body having rest mass $m_0$ moves along a trajectory $r_2:  R  \into R^3.$
Let $c$ denote the speed of light.
For any nonnegative function $k:R\into R$ define function $q:R \into R$
by the formula
\begin{equation*}
    q(t)=\sqrt{1-\frac{1}{(1+k(t)/(m_0c^2))^2}}\fa t\in   R  .
\end{equation*}

Then for any $t\in   R  $ the following two conditions are equivalent
\begin{itemize}
    \item The kinetic energy of the body $m_0$ on the
interval $(-\infty,t\rangle$ is bounded by $k(t).$
    \item The velocity $|v|$ of the body $m_0$ on the
interval $(-\infty,t\rangle$ is bounded by $c\,q(t).$
\end{itemize}
\end{pro}
\bigskip

\bp
    From Einstein's formula \cite{einstein2a}, p. 22,
    the kinetic energy of mass $m_0$ moving with the velocity $v$
    is given by the formula
    \begin{equation*}
        m_0c^2\left(\frac{1}{\sqrt{1-|v|^2/c^2}}-1\right).
    \end{equation*}
    Thus the condition
    \begin{equation*}
        m_0c^2\left(\frac{1}{\sqrt{1-|v(u)|^2/c^2}}-1\right)
        \le k(t)\fa u\le t,\,u\in   R
    \end{equation*}
    is equivalent to the condition
    \begin{equation*}
        |v(u)|\le c\,q(t)\fa u\le t,\,u\in   R  .
    \end{equation*}
    This completes the proof.
\ep
\bigskip

Notice that in the above proposition the quantity $q(t)<1$ for all $t\in R.$
\bigskip


\begin{thm}[Admissible trajectory is relativistic]
The notion of an admissible trajectory does not
depend on the Lorentzian frame.
\end{thm}

\bp
    Assume that we have two Lorentzian frames $S$ and $S'.$ Assume that
    the frame $S'$ moves away from frame $S$ with constant velocity $u.$
    Assume that $t\mapsto r_2(t)$ represents an admissible trajectory
    in the frame $S$ and a body with rest mass $m_0$ is moving along the
    trajectory.

    Without loss of generality we may assume
    that the frames $S$ and $S'$ are oriented so that the  transformation
    of the coordinates $y=(r,t)$ from $S$ to $S'$ is given by the formulas
    \begin{equation*}
    \begin{split}
    y'_1&=y_1\\
    y'_2&=y_2\\
    y'_3&=\gamma\,(y_3-uy_4)\\
    y'_4&=\gamma\,(y_4-uy_3)\\
    \end{split}
    \end{equation*}
    where $\gamma=(1-u^2)^{-1/2}$ and $y_4$ and $y'_4$ denote time in
    the respective frames. As before $c=1.$

    First of all notice that the time interval $(-\infty,\infty)$ maps
    onto itself from frame $S$ into $S'.$ Indeed we have
    \begin{equation*}
    \frac{dy'_4}{dy_4}=\gamma\ \frac{dy_4-u\,dy_3}{dy_4}
    =\gamma\,(1-uv)\ge\gamma\ (1-|u|)>0\fa t=y_4\in R.
    \end{equation*}
    Define function $g$ by the formula
    \begin{equation*}
    g(t)=y'_4(y_4)\fa t=y_4\in R.
    \end{equation*}
    From Cauchy's mean value theorem we have
    \begin{equation*}
    g(t)-g(0)=tg'(\theta)\ge t\,\gamma\ (1-|u|)\fa t>0.
    \end{equation*}
    Thus $y_4=g(t)\into \infty$ if $t\into \infty.$
    Similarly
    \begin{equation*}
    g(t)-g(0)=tg'(\theta)\le t\,\gamma\ (1-|u|)\fa t<0.
    \end{equation*}
    Thus $y_4=g(t)\into -\infty$ if $t\into -\infty.$
    Since any continuous function maps an interval onto an interval
    the function $g$ maps $R$ onto $R.$

    Introduce a function $f$ by the formula
    \begin{equation*}
        f(w)=\left(\frac{1}{\sqrt{1-w^2}}-1\right)\fa w\ge 0.
    \end{equation*}
    Notice that the function $f$ is nondecreasing and the kinetic
    energy of the mass $m_0$ moving along the trajectory
    can be represented as
    \begin{equation*}
    m_0f(|v|)
    \end{equation*}
    where $$v=\frac{dy_3}{dy_4}=\dot{r}_2$$
    is the  velocity of the body in the frame $S.$

    The velocity of the body in frame $S'$ is given by
    \begin{equation*}
    v'=\frac{dy'_3}{dy'_4}=\frac{dy_3-u\,dy_4}{dy_4-u\,dy_3}=\frac{v-u}{1-uv}.
    \end{equation*}
    Thus we have the estimate
    \begin{equation*}
    |v'|\le \frac{|v|+|u|}{1-|u|}\le \frac{q(t)+|u|}{1-|u|}
    \end{equation*}
    for all times in the initial interval $(-\infty, t\rangle.$
    The quantity  $q(t)$ denotes the velocity bound on the initial interval.
    Thus the velocity $v'$ is bounded on every initial interval $(-\infty,t'\rangle$
    in the frame $S'.$ Therefore its kinetic energy is bounded on every
    initial interval.

    Now let us consider the acceleration in the frame $S'.$ It can be
    expressed as
    \begin{equation*}
    \frac{dv'}{dy'_4}=\frac{\dot{v}(1-u^2)}{\gamma(1-uv)^3}
    \end{equation*}
    in terms of quantities in frame $S.$
    Thus on every initial interval $(-\infty,t\rangle$ we have the estimate
    \begin{equation*}
    \left|\frac{dv'}{dy'_4}\right|\le \frac{A(t)(1-u^2)}{\gamma(1-|u|)^3}
    \end{equation*}
    where $A(t)$ is the bound on the acceleration in the initial
    time interval $(-\infty,t\rangle$ in the frame $S.$
    Hence the acceleration in the frame $S'$ is bounded on every
    initial time interval $(-\infty,t'\rangle.$

    Therefore the trajectory in the frame $S'$ forms an admissible trajectory.
\ep

\section{Retarded time field}
\bigskip

Now consider any point $(r_1,t)$ in a fixed Lorentzian frame and let $(r_2(\tau),\tau)$
denote a point on the path of the point mass with the property that light
beam emitted from the trajectory will arrive at position $r_1$ at time $t.$

The time $\tau$ is called the {\bf retarded time.} It must satisfy the relation
\begin{equation*}
    |r_1-r_2(\tau)|^2-(t-\tau)^2=0,
\end{equation*}
which is preserved under Lorentzian transformations.

The following theorem establishes that the retarded time is
well defined as a function of the variables $(r_1,t)\in R^3\times R.$

\begin{thm}[The retarded time $\tau$ is unique and forms a continuous function]
\label{retarded time is unique}
Assume that we are given in a Lorentzian frame an
admissible trajectory $t\mapsto r_2(t).$
Then  for any point $r_1\in R^3$ and any time $t\in R$
there exists a unique number $\tau\le t$ such that
$$\tau=t-|r_1-r_2(\tau)|.$$

Moreover the map $(r_1,t)\mapsto \tau$ represents a locally Lipschitzian function on the
space $R^3\times R.$ Thus $\tau(r_1,t)$
is continuous on $R^3\times R.$
\end{thm}

\bp
    For fixed $r_1\in R^3$ and $t\in R$ introduce a function $f$
    by the formula
    \begin{equation*}
        f(s)=t-|r_1-r_2(s)|\fa s\le t.
    \end{equation*}
    The function $f$ is well defined and maps the closed interval $(-\infty,t\rangle$
    into itself. The function represents a contraction. Indeed
    \begin{equation}\label{tau is a contraction}
    \begin{split}
        |f(s)-f(\tilde{s})|&=\left|(t-|r_1-r_2(s)|)-(t-|r_1-r_2(\tilde{s})|)\right|
        \le |r_2(s)-r_2(\tilde{s})|\\
        &=|\int_{\tilde{s}}^s v(x)\,dx|\le v_1|s-\tilde{s}|\fa s,\tilde{s}\le t,
    \end{split}
    \end{equation}
    where $v_1=q(t)<c=1$ is the velocity bound corresponding to stopping time $t.$
    Therefore by Banach's contraction mapping theorem there exists one and only one
    solution of the equation $\tau=f(\tau).$

    Thus the map $(r_1,t)\mapsto \tau$ is well defined in our
    Lorentzian frame for all points  $(r_1,t)\in R^3\times R.$

    To prove that the function $\tau$ is locally Lipschitzian
    it suffices to prove that it is Lipschitzian on every open set
    of the form $R^3\times(-\infty,t_1).$  To this end take any two points $(r_1,t)$
    and $(\tilde{r}_1,\tilde{t})$ from the domain of $\tau$
    such that $t,\tilde{t}<t_1.$ Let $v_1<1$ denote the velocity bound corresponding
    to our trajectory on the interval $(-\infty,t_1).$

    To avoid unnecessarily
    complex notation denote by $\tau$ and $\tilde{\tau}$ the retarded times
    corresponding to the points
    $(r_1,t)$ and $(\tilde{r}_1,\tilde{t})$ respectively.
    We have
    \begin{equation*}
     \begin{split}
        |\tau -\tilde{\tau}|&=
        |f(\tau) -f(\tilde{\tau})|=
        \big|(t-|r_1-r_2(\tau)|)-(\tilde{t}-|\tilde{r}_1-r_2(\tilde{\tau})|)\big|\\
            & \le|t-\tilde{t}|+|r_1-\tilde{r}_1|+|r_2(\tau)-r_2(\tilde{\tau})|\\
            & =|t-\tilde{t}|+|r_1-\tilde{r}_1|+|\int_\tau^{\tilde{\tau}}\dot{r}_2(u)\,du|\\
            & \le|t-\tilde{t}|+|r_1-\tilde{r}_1|+v_1|\tau -\tilde{\tau}|.\\
     \end{split}
    \end{equation*}
    Taking the last term in  the above inequality onto the left side and dividing
    by $(1-v_1)$ both sides of the obtained inequality we get
    \begin{equation*}
        |\tau(r_1,t)-\tau(\tilde{r}_1,\tilde{t})|\le
        \frac{1}{1-v_1}(|t-\tilde{t}|+|r_1-\tilde{r}_1|)\fa
        (r_1,t),(\tilde{r}_1,\tilde{t})\in R^3\times (-\infty,t_1)
    \end{equation*}
    Thus the function $\tau$ is continuous on the entire space $R^3\times R.$
\ep

For a proof of Banach's contraction mapping theorem see,
for instance, Loomis and Sternberg
\cite{loomis} page 229.
\bigskip

\begin{thm}[An explicit formula for the retarded time function $\tau$]
\label{explicit formula for retarded time}
Assume that we are given in a Lorentzian frame an
admissible trajectory $t\mapsto r_2(t).$

Take any stopping time $t_1\in R$
and let $v_1=q(t_1)<c$ denote the corresponding velocity bound for $t\le t_1.$

Put $s_0(r_1,t)=0$ and define recursively the sequence
\begin{equation*}
    s_n(r_1,t)=f(s_{n-1}(r_1,t))\fa n=1,2,3,\dots;\text{ and }r_1\in R^3,\ t\le t_1,
\end{equation*}
where $f(s)=t-|r_1-r_2(t-s)|$ for all $s\le t.$

The retarded function $\tau$ is given by the formula
\begin{equation*}
    \tau(r_1,t)=\lim_n s_n(r_1,t) \fa  r_1\in R^3\text{ and }t\in R.
\end{equation*}
Moreover we have the following convenient estimate for the rate of convergence
\begin{equation*}
    |\tau(r_1,t)-s_n(r_1,t)|\le \frac{v_1^n}{1-v_1}\big|t-|r_1-r_2(t)|\big|\fa r_1\in R^3\text{ and }t\le t_1.
\end{equation*}
\end{thm}

\bp
The proof follows from Theorem 4.7 page 37 of Bogdan \cite{bogdan64} or
Theorem 9.1 on page 229
of Loomis and Sternberg \cite{loomis}.
\ep

\bigskip

\section{The fundamental fields associated with an
  admissible trajectory}
\bigskip

Now define  the delay function $T(r_1,t)=t-\tau(r_1,t)$ and notice that it satisfies
the equation
\begin{equation}\label{Equ for T}
    T=|r_1-r_2(t-T)|\fa t\in R\text{ and }r_1\in R^3.
\end{equation}

It is important to stress that in this paper we shall use the term {\bf field} as synonymous
with a function defined on a set of the space $R^4.$ The space $R^3\times R=R^4$ is
treated as a fixed Lorentzian frame. So we will use such expressions as scalar field,
vector field, tensor field, etc., to describe the functions taking values in corresponding
spaces. Thus the functions $\tau$ and $T$ represent continuous scalar fields defined
on the entire space $R^4.$

Since the function $T$ as difference of two continuous functions is continuous
the set
\begin{equation*}
 G=\set{(r_1,t)\in R^3\times R:\ T(r_1,t)>0}=T^{-1}(0,\infty)
\end{equation*}
as an inverse image of an open set by means of a continuous function is itself open.
The set $G$ consists of points that do not lie on the trajectory.

By assumption the trajectory $r_2$ has continuous derivatives $\dot{r}_2(t)=w(t)$
and $\dot{w}(t),$
so we can define the vector fields
$$r_{12}(r_1,t)=r_1-r_2(\tau(r_1,t)),\quad v(r_1,t)=\dot{r}_2(\tau(r_1,t))
,\quad a(r_1,t)=\dot{w}(\tau(r_1,t))$$
for all $(r_1,t)\in R^4.$
Introduce the unit vector field $e=r_{12}/T$ and
fields $u$ and $z$ by the formulas
$$
    u=\frac{1}{T}\qtext{and}z=\frac{1}{(1- \dotp{e}{v} )}\qtext{on}G.
$$
In the above $\dotp{e}{v}$ denotes the dot product of the vectors $e$ and $v.$
Since $|\dotp{e}{v}|\le |v|<c=1$ the vector field $z$ is well defined on the
set $G.$


\begin{defin}[Fundamental fields]
\label{Fundamental fields}
Assume that we are given in a Lorentzian frame an
admissible trajectory $t\mapsto r_2(t).$

Define the time derivative $w(t)=\dot{r}_2(t).$
The fields given by the formulas
\begin{equation*}
    \tau,\ T,\ r_{12}=r_1-r_2\circ \tau,\ v=w\circ \tau,\ a=\dot{w}\circ \tau\fa (r_1,t)\in G
\end{equation*}
and
\begin{equation*}
    u=1/T,\quad  e=u\, r_{12},\qtext{and} z\quad\fa (r_1,t)\in G
\end{equation*}
will be called the {\bf fundamental fields}
associated with the trajectory $r_2(t).$
The operation $\circ$ denotes here the composition of functions.
\end{defin}
\bigskip

The fundamental fields are continuous on their respective
domains.
This follows from the fact that composition of continuous functions yields
a continuous function. Thus all of them, for sure,
are continuous on the open set $G$ of
points that do not lie on the trajectory.

Analogous fields defined by similar formulas on an open set $G\subset R^4$
appear in the problems involving
plasma flows \cite{bogdan71} and \cite{bogdan72}, or more generally flows of matter.

\bigskip

We would like to stress here that the fundamental fields depend on the Lorentzian
frame, in which we consider the trajectory.
It is important to find expressions involving fundamental fields that
yield fields invariant under Lorentzian transformations.

Lorentz and Einstein \cite{einstein2a}, Part II, section 6, established that
fields satisfying Maxwell equations are invariant under Lorentzian transformations.
\bigskip

Our main goal is to prove that fields given by amended Feynman formulas and
fields obtained from Li\'{e}nard-Wiechert potentials satisfy
Maxwell equations. We shall do this by showing that these fields
are representable by means of fundamental fields and using the formulas
for partial derivatives of the fundamental fields prove that
such fields generate fields satisfying Maxwell equations.

The following theorem represents the main pillar of the entire
structure of the proof. Consider each of the formulas as
bricks from which the pillar is constructed. If any one
brick is rotten the whole structure of the proof will collapse.
This is just a gentle warning to an impatient reader not to
skip the computations involved.
\bigskip

Introduce operators $D=\frac{\partial}{\partial t}$ and
$D_i=\frac{\partial}{\partial x_{i}}$ for $i=1,2,3$ and
$\nabla=(D_1,D_2,D_3).$

Observe that $\delta_i$ in the following formulas denotes the i-th
unit vector of the standard base in $R^3$ that is
$\delta_1=(1,0,0),$ $\delta_2=(0,1,0),$ $\delta_3=(0,0,1).$


\begin{thm}[Partial derivatives of fundamental fields]
Assume that in some Lorentzian frame
we are given an   admissible trajectory $t\mapsto r_2(t).$
Define the time derivative $w(t)=\dot{r}_2(t).$
For partial derivatives with respect to coordinates of the vector $r_1$
we have the following identities on the set $G$
\begin{eqnarray}
\label{DiT}       D_iT&=&ze_i  \qtext{where} v=\dot{r}_2\circ\tau,\\
\label{Diu}       D_iu&=&-zu^2e_i  ,   \\
\label{Div}          D_iv&=&-e_iza \qtext{where} a=\dot{w}\circ\tau,\\
\label{Di tau}    D_i\tau&=&-ze_i,\\
\label{Die}         D_ie&=& -uze_ie+u\delta_i+uze_iv \qtext{where} \delta_i=(\delta_{ij}),\\
\label{Diz}          D_i z &=& -z^3e_i \langle e,a \rangle -uz^3e_i+ uz^2e_i+uz^2v_i+uz^3e_i \langle v,v \rangle \\
\label{grad T}\nabla T&=&ze  ,\\
\label{grad u}\nabla u&=&-zu^2e ,\\
\label{grad z} \nabla z&=&  -z^3 \langle e,a \rangle e-uz^3e+ uz^2e+uz^2v+uz^3 \langle v,v \rangle e.
\end{eqnarray}
and for the partial derivative with respect to time we have
\begin{eqnarray}
\label{DT}    DT&=&1-z,\\
\label{Du}    Du&=&zu^2-u^2,\\
\label{D tau}    D\tau&=&z,\\
\label{Dv}    Dv&=&za  \qtext{where} a=\dot{w}\circ\tau,\\
\label{De}    De&=&-u e+ u z e-u z v,\\
\label{Dz}    Dz&=&uz-2uz^2 +z^3 \langle e,a \rangle +uz^3-uz^3 \langle v,v \rangle .
\end{eqnarray}
Since the expression on the right side of each formula represents
a continuous function, the fundamental fields are at least of class $C^1$ on the set $G.$
Moreover if the trajectory is of class $C^\infty$ then also the fundamental
fields are of class $C^\infty$ on $G.$
\end{thm}

\bigskip

\bp
Proof of formula (\ref{DiT}):
Applying the operator $D_i$ to both sides of equation (\ref{Equ for T}) we get
\begin{equation*}
\begin{split}
    D_iT&=\langle e,D_i(r_1-r_2) \rangle
    =\langle e,(\delta_{ij}) \rangle -\langle e,v(\tau) \rangle D_i\tau\\
        =&e_i-\langle e,v(\tau) \rangle (-D_iT)=e_i+ \langle e,v \rangle D_iT
\end{split}
\end{equation*}
yielding formula (\ref{DiT}).

Formulas (\ref{Diu}), (\ref{grad T}), and
(\ref{grad u}) follow from formula (\ref{DiT}).

Proof of formula (\ref{Div}): $$D_iv=aD_i\tau=a(-D_iT)=-e_iza$$

Proof of formula (\ref{Di tau}): $$D_i\tau=D_i(t-T)=-D_iT=-ze_i$$

Proof of formula (\ref{Die}):
\begin{equation}
\begin{split}
    D_ie&=D_i[u(r_1-r_2)]=[D_iu]r_{12}+u[D_ir_1-D_ir_2]\\
    &=[D_iu]u^{-1}e+u[D_ir_1-D_ir_2]\\
    &=[-zu^2e_i]u^{-1}e+u\delta_i-u[D_i\tau]v\\
    &=-zue_ie+u\delta_i-u[-ze_i]v\\
    &=-zue_ie+u\delta_i+uze_iv\\
    &=-uze_ie+u\delta_i+uze_iv
\end{split}
\end{equation}

Proof of formula (\ref{Diz}):
\begin{equation}
\begin{split}
    D_i z&= D_i(1- \langle e,v \rangle )^{-1}=(-1)z^2(-D_i \langle e,v \rangle )\\
    &=z^2[ \langle e,D_iv \rangle + \langle v,-uze_ie+u\delta_i+uze_iv \rangle ]\\
    &=z^2[ \langle e,-e_iza \rangle + \langle v,-zue_ie+u\delta_i+uze_iv \rangle ]\\
    &=-z^3e_i \langle e,a \rangle -uz^3e_i
    \langle v,e \rangle +uz^2v_i+uz^3e_i \langle v,v \rangle \\
    &=-z^3e_i \langle e,a \rangle -uz^3e_i[1-z^{-1}]
    +uz^2v_i+uz^3e_i \langle v,v \rangle \\
    &=-z^3e_i \langle e,a \rangle -uz^3e_i+ uz^2e_i
    +uz^2v_i+uz^3e_i \langle v,v \rangle \\
\end{split}
\end{equation}

Formula (\ref{grad z}) follows from the formula (\ref{Diz}).
\bigskip

Proof of formula (\ref{DT}): Applying the operator
$D$ to both sides of the equation (\ref{Equ for T})
we get
\begin{equation*}
\begin{split}
    DT&=D|r_{12}|=(e,D[r_{12}])=-(e,Dr_2)\\
    &=- \langle e,v \rangle D\tau=- \langle e,v \rangle (1-DT)
    =- \langle e,v \rangle + \langle e,v \rangle DT.
\end{split}
\end{equation*}
The above yields
\begin{equation*}
\begin{split}
    DT&=\frac{- \langle e,v \rangle }{1- \langle e,v \rangle }=1-z.
\end{split}
\end{equation*}

Proof of formula (\ref{Du}): $$Du=DT^{-1}
=(-1)T^{-2}DT=(-1)u^2(1-z)=zu^2-u^2.$$

Proof of formula (\ref{D tau}): $$D\tau=D(t-T)=1-DT=1-(1-z)=z.$$

Proof of formula (\ref{Dv}): $$Dv=aD\tau=za.$$

Proof of formula (\ref{De}):
\begin{equation*}
\begin{split}
    De&=D[ur_{12}]=[Du]r_{12}+u[Dr_{12}]\\
    &=[zu^2-u^2]u^{-1}e-u[Dr_2]\\
    &=[zu-u]e-u[(D\tau)v]\\
    &=zue-ue-uzv=-ue+uze-uzv.
\end{split}
\end{equation*}

Proof of formula (\ref{Dz}):
\begin{equation*}
\begin{split}
    Dz&=D(1- \langle e,v \rangle )^{-1}
        =(1- \langle e,v \rangle )^{-2}D \langle e,v \rangle\\
    &=z^2 \langle De,v \rangle +z^2 \langle e,Dv \rangle \\
    &=z^2 \langle [uze-ue-uzv],v \rangle +z^2 \langle e,za \rangle \\
    &=uz^3 \langle e,v \rangle -uz^2 \langle e,v \rangle
    -uz^3 \langle v,v \rangle +z^3 \langle e,a \rangle \\
    &=uz^3[1-z^{-1}]-uz^2[1-z^{-1}]-uz^3 \langle v,v \rangle
    +z^3 \langle e,a \rangle \\
    &=uz^3-uz^2 -uz^2+uz-uz^3 \langle v,v \rangle +z^3 \langle e,a \rangle \\
    &=uz^3-2uz^2 +uz-uz^3 \langle v,v \rangle +z^3 \langle e,a \rangle \\
    &=uz-2uz^2 +z^3 \langle e,a \rangle +uz^3-uz^3 \langle v,v \rangle \\
\end{split}
\end{equation*}
\ep

\bigskip

\bigskip
\section{Electromagnetic potentials satisfying wave and  gauge equations\\
induce field satisfying Maxwell equations}
\bigskip

The following theorem represents the second main pillar of the structure.
Each step one should prove in detail to see that no heuristic is introduced.
The proof of this theorem one can trace back to the work of Lorentz \cite{lorentz1}.

\begin{thm}[Wave and  gauge imply Maxwell equations]
\label{wave-->Maxwell}
Let on an open set $G\subset R^4$
be given two scalar fields $\phi$ and $S$ and two vector fields $A$ and $J.$

Assume that the fields $\phi$ and $A$ have second partial derivatives with respect to the
coordinates of the point $(r_1,t)\in G$ and these derivatives are continuous on
the entire set $G.$

If these fields satisfy the following wave equations
\begin{equation*}
    \nabla^2\phi-\frac{\partial^2}{\partial t^2}\phi=-S,
    \quad\nabla^2A-\frac{\partial^2}{\partial t^2}A=-J,
\end{equation*}
and Lorentz gauge formula
\begin{equation*}
    \quad \nabla\cdot A+\frac{\partial}{\partial t}\phi=0
\end{equation*}
on the set $G,$
then the fields $E$ and $B$ defined by the formulas
\begin{equation*}
    E=-\nabla \phi - \frac{\partial}{\partial t} A\qtext{and}B
    =\nabla\times A\fa (r_1,t)\in G
\end{equation*}
will satisfy the following Maxwell equations
\begin{equation}\label{Maxwell equations for potentials}
\begin{split}
    &(a)\quad\nabla\cdot E=S,\ \quad(b)\quad\nabla\times E
    =- \frac{\partial}{\partial t} B,\ \quad\\
    &(c)\quad
    \nabla\cdot B=0,\ \quad(d)\quad\nabla\times B
    =\frac{\partial}{\partial t}E+J
\end{split}
\end{equation}
on the entire set $G.$
\end{thm}

\bigskip

\bp
    In this proof we will use the usual notation for dot product
    of two vectors $F\cdot H=\dotp{F}{H} $ to be close to the customary notation
    used in textbooks.
    The following proof represents a reversed argument presented in
    Feynman-Leighton-Sands \cite{feyn2}, vol. 2, chapter 18, section 6.
    Namely we show that potentials, that satisfy wave and Lorentz gauge equations,
    induce a pair of fields $E$ and $B$ satisfying Maxwell equations.
\bigskip

    Since for any vector field $F,$ which is twice continuously differentiable,
    we have $\nabla\cdot(\nabla\times F)=0,$ the vector field $B$ must satisfy
    Maxwell's equation $(c).$

    Since the differential operators $\nabla$ and $D$ commute, due to the
    assumption that the fields $\phi$ and $A$ are of class $C^2,$ and since from
    the Lorentz gauge formula follows that $\nabla\cdot A=-D\phi,$
    we must have
    \begin{equation*}
        \nabla\cdot E=\nabla\cdot(-\nabla\phi-DA)=-\nabla^2\phi-D\nabla\cdot A
        =-\nabla^2\phi+D^2\phi=S.
    \end{equation*}
    Therefore the equation $(a)$ is satisfied.

    Since $\nabla\times (\nabla h)=0$ for any scalar function,
    that is twice continuously differentiable on the set $G,$
    and for any such vector function $F$ defined on the open set $G$
    we have $\nabla\times DF=D\nabla\times F,$
    we get
    \begin{equation*}
        \nabla\times E=\nabla\times(-\nabla\phi-DA)=-D\nabla\times A=-DB
    \end{equation*}
    and this means that Maxwell's equation $(b)$ is satisfied.

    Finally for the last equation, Maxwell's equation $(d)$, from the
    algebraic identity
    \begin{equation*}
        \nabla\times(\nabla\times A)=\nabla(\nabla\cdot A)-\nabla^2 A
    \end{equation*}
    and Lorentz gauge equation and the wave equation
    $$\nabla\cdot A+D\phi=0,\quad\nabla^2A-D^2A=-J$$
    we get
    \begin{equation*}
        \begin{split}
            \nabla\times B - D\,E&=\nabla\times(\nabla\times A)-D(-\nabla\phi-D\,A)\\
                &=\nabla(\nabla\cdot A)-\nabla^2 A + D\nabla\phi + D^2A\\
            &=\nabla[(\nabla\cdot A)+D\phi]-(\nabla^2A-D^2A)=J.
        \end{split}
\end{equation*}
\ep
\bigskip

Notice that in the above theorem the field $S$ corresponds
to the field $\varrho\,$ of charge density
and $J$ to the field $j$ of intensity of charge flow. Since
we redefined the electrostatic constant $\epsilon_0$ to satisfy the condition
$4\pi \epsilon_0=1$ our new unit of charge became equal to $4\pi$ [coulomb].
\bigskip

It is worthwhile to mention also
the following proposition.
\bigskip

\begin{pro}[Equation of continuity for flow of charge]
If the electromagnetic potentials $\phi$ and $A$ have continuous
partial derivatives of order 3 on an open set $G\subset R^4,$
then from the Lorentz gauge equation
$$
    D\phi+\nabla\cdot A=0\fa (r_1,t)\in G
$$
follows the equation
$$
    DS+\nabla\cdot J=0\fa (r_1,t)\in G
$$
of continuity for charge flow.
\end{pro}
\bp
    Indeed introduce the D'Alembertian operator
    $$\Box^2=D^2-\nabla^2.$$ From our assumptions follows
    the commutativity of the differential operators
    $$D,\quad \nabla,\quad \Box^2$$  when acted on fields $\phi$ and $A,$
    respectively.
    Hence we have
    $$\Box^2D=D\,\Box^2\qtext{and}\Box^2\nabla=\nabla\,\Box^2.$$
    The above implies
    $$DS+\nabla\cdot J=D\,\Box^2\phi+\nabla\cdot \Box^2 A
    =\Box^2(D\phi+\nabla\cdot A)=0$$
    and this is the equation of continuity of flow of charge.
\ep

\bigskip
\section{Li{\'e}nard-Wiechert potentials and wave and gauge equations}
\bigskip

In this section we will deal only with Li\'{e}nard-Wiechert potentials
$$
    \phi=uz\qtext{and}A=uzv.
$$

The author in the proof of the following theorem used DOS
based interactive computer software called
DERIVE allowing to input user data files and perform on them interactive
operations. More details about DERIVE and
actual programs used in computation we shall present
in the last section of the paper.
\bigskip

Any program, that  can symbolically manipulate polynomials of an arbitrary
number of variables, can be used to verify the proof.
For a reader, who would like
personally to verify the proof by hand or another software system,
we present here just the key intermediate
formulas as computational mile stones.

In the following formulas the vector field $\dot{a}$ denotes the
composition of the function $\frac{d^3}{dt^3}r_2(t)$ with the
retarded time field $\tau.$

\begin{thm}[Homogenous wave and gauge equations are satisfied]
\label{wave equations}
For any   admissible trajectory $r_2(t)$ of class $C^3$ the two
homogenous wave equations
\begin{equation*}
    (\nabla^2-D^2)[uz]=0,\quad(\nabla^2-D^2)[uzv]= 0,
\end{equation*}
 and Lorentz gauge equation
\begin{equation*}
    \nabla\cdot[uzv]+D[uz]=0
\end{equation*}
are satisfied on the entire set  $G=\set{(r_1,t)\in R^4:\ T(r_1,t)>0}.$
\end{thm}

\bigskip

\bp
We have the following identities

\begin{equation*}
\begin{split}
  \nabla^2[uz]=D^2[uz]=&
+  u z^4 \langle e,\dot{a} \rangle
+  u^2 z^3 \langle e,a \rangle
+3 u z^5 \langle e,a \rangle^2
+3 u^3 z^3\\
&+3 u^3 z^5
+3 u^3 z^5 \langle v,v \rangle ^2
+6 u^2 z^5 \langle e,a \rangle
+6 u^3 z^4 \langle v,v \rangle\\
&-3 u^2 z^4 \langle v,a \rangle
-6 u^2 z^4 \langle e,a \rangle
-6 u^2 z^5 \langle v,v \rangle  \langle e,a \rangle
-6 u^3 z^4\\
&-6 u^3 z^5 \langle v,v \rangle
-u^3 z^3 \langle v,v \rangle
\end{split}
\end{equation*}
Thus the first wave equation takes the form
\begin{equation*}
    \nabla^2[uz]-D^2[uz]=0.
\end{equation*}

\bigskip

Similarly for the second wave equation we have the formulas
\begin{equation*}
\begin{split}
&\nabla^2[uzv]=D^2[uzv]=\\
&=
+  u z^3 \dot{a}
+  u z^4 \langle e,\dot{a}  \rangle  v
+  u^2 z^2 a
+  u^2 z^3 \langle e,a \rangle  v\\
&+3 u z^4 \langle e,a \rangle  a
+3 u z^5 \langle e,a \rangle ^2 v
+3 u^2 z^4 a
+3 u^3 z^3 v\\
&+3 u^3 z^5 \langle v,v \rangle ^2 v
+3 u^3 z^5 v
+6 u^2 z^5 \langle e,a \rangle  v
+6 u^3 z^4 \langle v,v \rangle  v\\
&-  u^3 z^3 \langle v,v \rangle  v
-3 u^2 z^4 \langle v,a \rangle  v
-3 u^2 z^4 \langle v,v \rangle  a
-4 u^2 z^3 a\\
&-6 u^2 z^4 \langle e,a \rangle  v
-6 u^2 z^5 \langle v,v \rangle  \langle e,a \rangle  v
-6 u^3 z^4 v
-6 u^3 z^5 \langle v,v \rangle  v.
\end{split}
\end{equation*}

Computing terms of Lorentz gauge equation we get
\begin{equation*}
\nabla\cdot[uzv]= \langle \nabla,uzv \rangle=
+u^2 z^3 \langle v,v \rangle -u^2 z^3+u^2 z^2-u z^3 \langle e,a \rangle
\end{equation*}
and
\begin{equation*}
 D[uz]=
-u^2 z^3 \langle v,v \rangle +u^2 z^3-u^2 z^2+u z^3 \langle e,a \rangle.
\end{equation*}
Hence
\begin{equation*}
 \langle \nabla,uzv \rangle+ D[uz]=0.
\end{equation*}
\ep
\bigskip

Li\'{e}nard-Wiechert
potentials generate a solution to a homogenous system of Maxwell equations.
Indeed from previous theorems we get the following corollary.

\begin{cor}[Li\'{e}nard-Wiechert potentials and Maxwell equations]
Let $r_2(t)$ be any   admissible trajectory and
let $G$ denote the open set of points that do not lie on the trajectory, that is
$$G=\set{(r_1,t)\in R^4:\ T(r_1,t)>0}.$$
Define the fields by formulas
\begin{equation*}
 \phi=uz,\quad A=uzv,\quad S=0,\quad J= 0\fa (r_1,t)\in G.
\end{equation*}
Then the fields $E$ and $B$ defined by the equalities
\begin{equation*}
    E=-\nabla \phi - \frac{\partial}{\partial t} A\qtext{and}B
    =\nabla\times A \fa (r_1,t)\in G
\end{equation*}
will satisfy the following homogenous Maxwell equations
\begin{equation*}
    \nabla\cdot E=0,\quad\nabla\times E
    + \frac{\partial}{\partial t} B=0,\quad
    \nabla\cdot B=0,\quad\nabla\times B
    -\frac{\partial}{\partial t}E=0
\end{equation*}
on the set $G.$
\end{cor}
\bigskip

\section{Proof of the amended Feynman's formulas}
\bigskip

As before let $r_2(t)$ be any   admissible trajectory of class $C^3$
in some Lorentzian frame
and let $$G=\set{(r_1,t)\in R^4:\ T(r_1,t)>0}$$ be the set of points not lying
on the trajectory.
Let $\phi=uz$ and $A=uzv$ denote the  Li\'{e}nard-Wiechert potentials
corresponding to the trajectory $r_2.$
\bigskip

We remind the reader that all intensities of the considered fields are
per unit of charge, thus if we are dealing with a charge $q$ the
corresponding potentials will take the form $\phi=quz$ and $A=quzv$
and similar adjustments should be made with the fields $E,\ B.$
\bigskip

Feynman using his formula for the electromagnetic potentials,
see  vol. 2, Feynman-Leighton-Sands \cite{feyn2}, chapter 15, page 15.15,  had derived
on page 21.9, of chapter 21, a  formula for
electromagnetic potential of a moving point charge. A formula,
which did not agree with the Li\'{e}nard-Wiechert formulas.

Nevertheless on page 21.11 in the last sentence he suggested
that perhaps the reader may take his word for it, that  Li\'{e}nard-Wiechert
potentials yield the same field as his formulas in the case of a moving
point charge.
In the footnote he
mentioned that he has done that, but it took him a lot of time and paper.

This little footnote provided the author with the main impetus
to try to verify the computations. Feynman did not leave any
hints concerning possible simplifications of the computations
or that quantities involved should be considered as fields.

The author taking Feynman's suggestion performed by hand, step by step,
computations presented in the following proof.
These computations have provided an essential
hint how to develop and to structure the proofs of the theorems presented in
this paper.
\bigskip

Thus the credit has to be given with gratitude to Feynman for discovery of this gem stone
of {\em Einstein's Special Theory of Relativity} and consequently of {\em Maxwell's
Theory of Electrodynamics.}
\bigskip

To verify
in a more efficient way that the electric fields $E$ and $E_f$
coincide, we provide a computer program in the last section of this paper.
However to show that the magnetic fields $B$ and $B_f$ are equal
a direct computation seems to be more efficient.

To Feynman are due immense thanks and gratitude beyond words!
Here are the computations as they might have
looked under Feynman's hand.

\begin{thm}[Equality of amended Feynman and Li\'{e}nard-Wiechert fields]
Assume that in some Lorentzian frame we are given
an admissible trajectory $r_2(t)$ of class $C^3.$

Let $$G=\set{(r_1,t)\in R^4:\ T(r_1,t)>0}$$ denote the set of points not lying
on the trajectory.

Let $u$ and $e$ denote the fundamental fields (\ref{Fundamental fields})
generated by the trajectory $r_2(t).$

Define fields
\begin{equation*}
\begin{split}
    E\ \,&=-\nabla \phi-DA\qtext{and}B=\nabla\times A,\\
    E_f&=u^2e+u^{-1}D(u^2e)+D^2e\qtext{and}B_f=e\times E_f,\\
\end{split}
\end{equation*}
Then we have the equalities $E=E_f$ and $B=B_f$ on the set $G.$
\end{thm}
\bigskip

\bp
From the formulas (\ref{grad z}) and  (\ref{grad u}) we get
\begin{equation*}\label{grad phi}
    \begin{split}
    \nabla\phi &=\nabla ( z u)=u\nabla z + z\nabla u\\
    &=u[-z^3 \langle e,a \rangle e-uz^3e+ uz^2e+uz^2v+uz^3 \langle v,v \rangle e] + z[-zu^2e]\\
    &=[-uz^3 \langle e,a \rangle e-u^2z^3e+ u^2z^2e+u^2z^2v+u^2z^3 \langle v,v \rangle e] -u^2z^2e\\
    &=-uz^3 \langle e,a \rangle e+u^2z^2v-u^2z^3e+u^2z^3 \langle v,v \rangle e\\
    \end{split}
\end{equation*}
and
\begin{equation*}
    \begin{split}
    D[u v]&=[Du]v+uDv=[zu^2-u^2]v+uza\\
        &=u^{2}zv-u^2v+uza\\
        &=uza-u^2v+u^{2}zv\\
    \end{split}
\end{equation*}

\begin{equation*}\label{DA}
    \begin{split}
     DA&= D[ z u v]=[D z]uv+ z D[uv]\\
    &=[D z]uv+ z D[uv]\\
    &=[uz-2uz^2 +z^3 \langle e,a \rangle +uz^3-uz^3 \langle v,v \rangle ]uv\\
    &\quad + z [uza-u^2v+u^{2}zv]\\
    &=u^2zv-2u^2z^2v +uz^3 \langle e,a \rangle v+u^2z^3v\\
    &\quad -u^2z^3 \langle v,v \rangle v +uz^2a-u^2zv+u^{2}z^2v\\
    &= uz^2a-u^2z^2v +uz^3 \langle e,a \rangle v+u^2z^3v-u^2z^3 \langle v,v \rangle v   \\
    \end{split}
\end{equation*}
thus
\begin{equation*}\label{formula4E}
\begin{split}
    E&=-\nabla \phi -DA\\
    &=-[-uz^3 \langle e,a \rangle e+u^2z^2v-u^2z^3e+u^2z^3 \langle v,v \rangle e]\\
    &\qquad -[ uz^2a-u^2z^2v +uz^3 \langle e,a \rangle v+u^2z^3v-u^2z^3 \langle v,v \rangle v]\\
    &=-[ uz^2a  -uz^3 \langle e,a \rangle e+uz^3 \langle e,a \rangle v \\
    &\quad  -u^2z^3e+u^2z^3 \langle v,v \rangle e+u^2z^3v-u^2z^3 \langle v,v \rangle v]\\
\end{split}
\end{equation*}
Now consider the second term of the expression $E_f.$ We have
\begin{equation*}\label{second}
\begin{split}
    u^{-1}D[{u^{2}}{e}]&=u^{-1}[Du^{2}]e+uDe\\
        &=u^{-1}2u[Du]e+u[-ue+uze-uzv]\\
        &=2[u^2z-u^2]e-u^2e+u^2ze-u^2zv\\
        &=2u^2ze-2u^2e-u^2e+u^2ze-u^2zv\\
        &=3u^2ze-3u^2e-u^2zv\\
        &=-3u^2e+3u^2ze-u^2zv\\
        &=2u^{2}(z-1)e+u[uze-ue-uzv]\\
        &=2u^{2}ze-2u^{2}e+u^{2}ze-u^{2}e-u^{2}zv\\
        &=3u^{2}ze-3u^{2}e-u^{2}zv\\
\end{split}
\end{equation*}
For the third term of the expression $E_f$ we get
\begin{equation*}\label{third}
\begin{split}
    D^2e&=D[De]=D[-ue+uze-uzv]=D\{u[z(e-v)-e]\}\\
    &=\{Du\}[z(e-v)-e]+uD[z(e-v)-e]\\
    &=\{u^{2}z-u^2\}[ze-zv-e]\\
    &\quad +u[(Dz)(e-v)+zD(e-v)-De]\\
    &=       2u^2e  -3u^2ze+u^2zv-uz^2a   +uz^3 \langle e,a \rangle e -uz^3 \langle e,a \rangle v\\
    &\quad   +u^2z^3e-u^2z^3 \langle v,v \rangle e-u^2z^3v+u^2z^3 \langle v,v \rangle v\\
\end{split}
\end{equation*}
Compute the field $E_f$
\begin{equation*}\label{formula4F}
\begin{split}
    E_f&=u^2e+u^{-1}D(u^2e)+D^2e\\
    &=u^2e+[-3u^2e+3u^2ze-u^2zv]\\
    &+[2u^2e  -3u^2ze+u^2zv-uz^2a   +uz^3 \langle e,a \rangle e -uz^3 \langle e,a \rangle v\\
    &\qquad   +u^2z^3e-u^2z^3 \langle v,v \rangle e-u^2z^3v+u^2z^3 \langle v,v \rangle v\\
    &= -uz^2a   +uz^3 \langle e,a \rangle e -uz^3 \langle e,a \rangle v \\
    &\quad  +u^2z^3e-u^2z^3 \langle v,v \rangle e-u^2z^3v+u^2z^3 \langle v,v \rangle v\\
\end{split}
\end{equation*}
and compare with
\begin{equation*}
\begin{split}
    E&=-[ uz^2a  -uz^3 \langle e,a \rangle e+uz^3 \langle e,a \rangle v  \\
    &\quad -u^2z^3e+u^2z^3 \langle v,v \rangle e+u^2z^3v-u^2z^3 \langle v,v \rangle v].
\end{split}
\end{equation*}
Now the magnetic field from Feynman's formula is
\begin{equation*}
 \begin{split}
    B_f&=e\times E_f=e\times [-uz^2a  -uz^3 \langle e,a \rangle v
        -u^2z^3v+u^2z^3 \langle v,v \rangle v]\\
        &=-uz^2\ e\times a  -uz^3 \langle e,a \rangle \ e\times v
        -u^2z^3\ e\times v+u^2z^3 \langle v,v \rangle \ e\times v\\
\end{split}
\end{equation*}
and since
\begin{equation*}
\begin{split}
     \nabla [uz] &= -uz^3 \langle e,a \rangle e+u^2z^2v-u^2z^3e+u^2z^3 \langle v,v \rangle e,\\
     \nabla\times v&=-z\ e\times a,
\end{split}
\end{equation*}
we have from Li\'{e}nard-Wiechert potentials that
\begin{equation*}
     \begin{split}
        B&=\nabla\times A=\nabla\times [uzv]=[uz]\ \nabla\times v+\nabla[uz]\times v\\
        &=[uz](-z\ e\times a)+(-uz^3 \langle e,a \rangle e+u^2z^2v\\
        &\quad -u^2z^3e+u^2z^3 \langle v,v \rangle e)\times v\\
        &=-uz^2\ e\times a -uz^3 \langle e,a \rangle \ e \times v-u^2z^3\ e \times v
            +u^2z^3 \langle v,v \rangle \ e \times v\\
    \end{split}
\end{equation*}
Clearly we have the equality for both fields $$E=E_f\qtext{ and }B=B_f.$$
{\bf Eureka!  A salute is due to Feynman!}
\ep

Now let us collect the formulas representing the fields $E$ and $B$
by means of the fundamental fields (\ref{Fundamental fields})
into one convenient statement.
These formulas can be very useful for computations especially with the help
of a digital computer.

Notice that all the fundamental fields generated by the admissible trajectory
are defined by
means of algebraic operations and operation of composition of functions,
with the exception of the retarded time field $\tau,$ but Banach's fixed point
theorem (\ref{explicit formula for retarded time}) provides a fast converging algorithm
for computing this function.
\bigskip

\begin{pro}[Representation of amended Feynman's fields by fundamental fields]
\label{Representation of Feynman's fields}
Assume that in some Lorentzian frame we are given
an admissible trajectory $r_2(t)$ of class $C^3$ of a moving point mass $m_0.$
Let $$G=\set{(r_1,t)\in R^4:\ T(r_1,t)>0}$$ denote the set of points not lying
on the trajectory.
Define fields
\begin{equation*}
\begin{split}
    E&=u^2e+u^{-1}D(u^2e)+D^2e\qtext{and}B=e\times E,\\
\end{split}
\end{equation*}
Then the fields $E$ and $B$ have the following representation
in terms of the fundamental fields (\ref{Fundamental fields})
generated by the trajectory $r_2(t).$

\begin{equation*}
\begin{split}
    E    &= -uz^2a   +uz^3 \langle e,a \rangle e -uz^3 \langle e,a \rangle v \\
    &\quad  +u^2z^3e-u^2z^3 \langle v,v \rangle e-u^2z^3v+u^2z^3 \langle v,v \rangle v,\\
    B&=-uz^2\ e\times a  -uz^3 \langle e,a \rangle \ e\times v
        -u^2z^3\ e\times v+u^2z^3 \langle v,v \rangle \ e\times v,\\
\end{split}
\end{equation*}
\end{pro}
\bigskip

\section{Modified theory of generalized functions}
\bigskip

Now  let us prove that any pair $(E,B)$ of fields of class $C^2$
on an open set $G$ in $R^4$
that satisfies the homogenous system of Maxwell equations
must satisfy the homogenous wave equations on $G.$

It will be convenient here to use the theory of generalized functions, called
also distributions,
originally introduced heuristically by Dirac in his works
on Quantum Mechanics and put on precise mathematical footing by
L. Schwartz \cite{schwartz} and Gelfand
and Shilov \cite{gelfand1}. We do not introduce here
any notions from the theory of Lebesgue-Bochner integral.
We base the proofs only on elementary properties
of continuous and differentiable functions that can
be found in any book on Calculus.

We will need to make a few modifications
for the sake of our problem. We need distributions
defined over any open set $G\subset R^4$ and
having their values in the vector space $R^k.$

Let $\D_k$ denote the space $C^\infty_0(G,R^k)$
of infinitely differentiable functions having compact supports contained in $G$ and
values in $R^k.$

On the space $\D_k$ we introduce the usual sequential topology. A sequence
of functions $g_n\in \D_k$ converges to a function $g\in \D_k$ if it
converges uniformly together with all its partial derivatives $D^\alpha g$,
on every compact subset $K$ of $G,$ to the function $g,$ that is for every $\alpha $
the following sequence converges uniformly on $ K $
$$
    D^\alpha g_n(x)\into D^\alpha g(x)
$$
where $\alpha=(k_1,\cdots,k_4)$ denotes the multi-index of a partial derivative
$$D^\alpha=D_1^{k_1}\cdots D_4^{k_4}$$
with
$k_j=0,1,2,\ldots$ and $D_j=\frac{\partial}{\partial x_j}$ where $j=1,2,3,4.$

\begin{defin}[Generalized vector valued functions]
\label{Generalized vector valued functions}
Let $\D'_k$ denote the space of all linear continuous real functionals on $\D_k.$
Convergence in $\D'_k$ will be understood as pointwise convergence. The space
$\D'_k$ will be called the space of {\bf generalized vector valued functions} or
the space of vector distributions.
\end{defin}

We shall use the following equivalent notation for such a functional
\begin{equation}\label{integral formula for distribution}
    f(g)=\dotp{f}{g}=\int_G f(x)\cdot g(x)\,dx\fa g\in \D_k.
\end{equation}
In the above  $y\cdot x$
denotes the dot product, also called the scalar product, of two vectors $y,x\in R^k.$

Any continuous function $f$ on the set $G$ generates by means of the above integral
formula a vector distribution.

Notice that
the space $\D'_k$ is linearly and topologically isomorphic with the Cartesian
product of $k$ copies of the space $\D'_1$
$$
    \D'_1\times\cdots\times\D'_1=(\D'_1)^k.
$$

Indeed, if for every $m=1,\dots,k$ we define maps $P_m:R\into R^k$ by the condition
\begin{equation*}
    P_m(t)=x\Leftrightarrow \{x_m=t\text{ and }x_j=0\text{ if }j\not=m\},
\end{equation*}
where $x=(x_1,\ldots,x_k)\in R^k,$
then the functionals
\begin{equation*}
    f_m(g)=\int_G f(x)\cdot P_m(g(x))\,dx\fa g\in \D_1
\end{equation*}
are well defined and represent elements of the space $\D'_1.$
Thus the element
\begin{equation*}
 (f_1,\cdots,f_k)\in (\D'_1)^k.
\end{equation*}

Conversely define maps $P'_m:R^k\into R$ for $m=1,\ldots,k$ by the formula
\begin{equation*}
    P'_m(x)=x_m\fa x\in R^k,\ m=1,\ldots,k.
\end{equation*}
Then the  formula
\begin{equation*}
    f(g)=\sum_m\int_G f_m P'_m(g(x))\,dx\fa g\in \D_k
\end{equation*}
yields a linear continuous functional on the space $\D_k.$
It is easy to verify that the transformation $Q:\D'_k\into (\D'_1)^k$defined by
\begin{equation*}
    f\mapsto (f_1,\ldots,f_k)
\end{equation*}
is indeed a linear and topological isomorphism of the two spaces.
\bigskip

Now notice the following fact.

\begin{pro}[Imbedding of continuous functions into $\D'_k$ is one-to-one]
Given two continuous functions $f_1$ and $f_2$ on the open set $G$
such that they generate the same
vector distribution, that is
\begin{equation*}
    \int_G f_1(x)g(x)\,dx=\int_G f_2(x)g(x)\,dx\fa g\in \D_k,
\end{equation*}
then they coincide
\begin{equation*}
    f_1(x)=f_2(x)\fa x\in G.
\end{equation*}
\end{pro}
\bigskip

\bp
Indeed, from linearity of the map $f\mapsto \dotp{f}{g}$ and the previous isomorphism
it is sufficient to prove that for any real continuous function $f$
such that
\begin{equation*}
    \int_G f(x)g(x)\,dx=0\fa g\in \D_1
\end{equation*}
follows that $f=0$ on $G.$
\bigskip

Assuming that this is not true
then at some point $x_0\in G$ we have $f(x_0)\neq 0.$ We may assume without loss of
generality that $2\delta=f(x_0)>0$ otherwise we would consider the function $-f.$
From continuity of $f$ follows that there is a rectangular
neighborhood $V\subset G$ of $x_0$ such that
\begin{equation*}
    f(x)\ge \delta \fa x\in V.
\end{equation*}

There exists a nonnegative function $g$ of class $C^{\infty}$ with
support in the set $V$ with integral $\int_G g(x)\,dx=1.$ Thus
for such a function we would get
\begin{equation*}
    \int_G f(x)g(x)\,dx=\int_V f(x)g(x)\,dx\ge \int_V \delta g(x)\,dx=\delta>0.
\end{equation*}
A contradiction. So all we have to do is to show that there exists
a function $g$ having the above properties. To this end
consider a nonnegative function $g_0(t)$ defined by the formula
\begin{equation*}
    g_0(t)=\alpha e^{-1/(1-t^2)}\text{ if }|t|<1;\qtext{and}g_0(t)=0\text{ if }|t|\ge 1,
\end{equation*}
where the constant is selected so that $\int_{-1}^1g_0(t)\,dt=1.$
It follows from the above formula that the function $g_0$ is
infinitely differentiable at every point except perhaps at $t=+1$ or $t=-1.$
One can prove that at these points the one-sided derivatives exist and
they are equal. Thus the function $g_0$ is of class $C^\infty.$

Now consider the function
\begin{equation*}
    g_1(x)=g_0(x_1)g_0(x_3)g_0(x_2)g_0(x_4)\fa x=(x_1,x_2,x_3,x_4)\in R^4
\end{equation*}
with the integral, over the cube in $R^4$ representing its support, equal to 1.
For the fixed $x_0\in G$ and sufficiently large $n$ we see that the function
given by the formula
\begin{equation*}
    g_n(x)=n^4g(n(x-x_0))\fa x\in R^4
\end{equation*}
will satisfy our requirements.
\ep

Any linear continuous operator $H :\D_k\into\D_k$ generates a linear continuous
dual operator $H ':\D'_k\into\D'_k$ by the formula
\begin{equation*}
    \dotp{H 'f}{g}=\dotp{f}{H \, g}\fa g\in \D_k.
\end{equation*}

The dual operator corresponding to scalar multiplication $g\mapsto \lambda\, g$
is scalar multiplication  $f\mapsto \lambda\, f$ as follows from
the above definition.

\begin{defin}[Generalized differential operator]
\label{generalized partial derivative}
By a {\bf generalized partial derivative} $D_i=\frac{\partial}{\partial x_i}$
acting onto the m-th component of $f\in \D'_k$
\begin{equation*}
    D_{i,m}(f_1,\ldots, f_k)=(f_1,\ldots,D_i f_m,\ldots,f_k)
\end{equation*}
we shall understand the dual operator to the operator acting onto
the m-th component of $g\in\D_k$ by the formula
$$
    (-1)D_{i,m}(g_1,\ldots, g_k)=(g_1,\ldots,(-1)D_i g_m,\ldots,g_k).
$$
\end{defin}
\bigskip

In the case when the set $G$ represents a Cartesian product of open
bounded intervals and the vector function $f$ is continuous together with
$D_if_m,$
the above formula can be easily verified through iterated integral and
integration by parts.
In this case the ordinary partial derivative
will produce the derivative in the sense of distributions.
So it is natural to extend this property to vector distributions.
\bigskip

\begin{defin}[Weak and strong partial derivatives]
\label{def-weak and strong partial derivatives}
Now every distribution and in particular every continuous function is
differentiable in the sense of distributions. If it happens that the
distributional partial derivative is representable by means of a continuous
function, such a function is called a {\bf weak derivative.}
If a vector function
has a continuous partial derivative such a derivative is called
a {\bf strong derivative.}
\end{defin}
\bigskip

It is not obvious that weak and strong derivatives coincide in
the general case of an arbitrary open set $G$ in $R^4.$
This calls for the following theorem.
\bigskip

\begin{thm}[Weak and strong partial derivatives coincide]
\label{weak and strong derivatives coincide}
Assume that $f=(f_1,\ldots,f_k)$ is a vector-valued function on
an open set $G\subset R^4$ and that its $m$-th component
has a continuous partial derivative $D_if_m.$

Then this derivative coincides with the derivative in the sense of distribution,
that is the weak derivative coincides with the strong one.
\end{thm}
\bigskip

\bp
To prove this fact for general open sets in $R^4$
notice that for every point $x\in G$ there exists a neighborhood $V(x)\subset G$ in the form
of the Cartesian product of open intervals.
Restricting our test functions
to functions with support in $V(x)$ will yield that on such a domain
the weak and strong partial derivatives coincide.

Since every open set in
$R^4$ is a union of a countable number of compact sets, from the collection
$V(x)\,(x\in G)$ one can extract a locally finite countable cover of the set $G.$

Now using partition of unity theorem,
(for reference concerning this theorem
see for instance Gelfand and Shilov \cite{gelfand1},
vol. 1, Appendix to Chapter 1, Section 2,)
we can prove this theorem for any open set $G\subset R^4.$
\ep
\bigskip

Now let $\A$ denote the algebra of all linear transformations $H:\D'_k\into\D'_k.$
The multiplication of two transformations $H$ and $S$ in the algebra $\A$ is defined
as composition $H\circ S.$

Since any two partial derivatives commute and they commute with scalar multiplication
operators, the above definition \ref{generalized partial derivative}
permits one to represent any polynomial of the
operators $D_{i,m}$ as a dual of some operator acting in the space $\D_k.$
It follows from the above definition that any operator $H,$
generated by a homogenous quadratic polynomial $p\,,$ is formally self
dual
\begin{equation*}
    \dotp{Hf}{g}=\dotp{f}{Hg}\fa g\in\D_k.
\end{equation*}

Now let us consider a particular space of interest $\E=\D'_3\times \D'_3.$

 Let $\A_0$ denote the algebra of all linear transformations
from $\D'_3$ into itself with multiplication defined again as a composition
of transformations.

\begin{pro}[Matrix representation of operators on $\E$]
\label{Matrix representation of operators}
A transformation $H$ belongs to the algebra $\A$ if and only if there exists
a $2\times 2$ matrix such that
\begin{equation}\label{Matrix representation}
    H(f)=\begin{bmatrix}
      H_{11},H_{12} \\
      H_{21},H_{22} \\
    \end{bmatrix}
    \begin{bmatrix}
      f_1 \\
      f_2 \\
    \end{bmatrix}
    \fa f=   \begin{bmatrix}
      f_1 \\
      f_2 \\
    \end{bmatrix}
    \in \E,
\end{equation}
where transformations $H_{ij}$ belong to algebra $\A_0.$
This correspondence establishes isomorphism between the algebra $\A$
and the algebra of all $2\times 2$ matrices with elements in the
algebra $\A_0.$
\end{pro}

\bp
    The proof is evident and we leave it to the reader.
\ep


A transformation $H$ with matrix of the form
$a_{ij}I,$ where $a_{ij}\in R$ and $I$ is the identity map in $\A_0,$
will be called scalar transformation. A scalar transformation $H$ can be simply
represented as
\begin{equation*}
    H(f)=
    \begin{bmatrix}
      a_{11} & a_{12} \\
      a_{21} & a_{22} \\
    \end{bmatrix}
    \begin{bmatrix}
      f_1 \\
      f_2 \\
    \end{bmatrix}
    \fa f\in \E.
\end{equation*}

Notice that the identity transformation, the unit element $I$ in the algebra $\A,$
can be represented as a {\bf scalar transformation}
with matrix
\begin{equation*}
    \begin{bmatrix}
      1 & 0 \\
      0 & 1 \\
    \end{bmatrix}.
\end{equation*}

Of a particular interest are the following two scalar transformations that are inverse
to each other
\begin{equation*}
    J=
    \begin{bmatrix}
      0 & -1 \\
      1 &0 \\
    \end{bmatrix}
    \qtext{and}
    J^{-1}=
    \begin{bmatrix}
      0 & 1 \\
      -1 &0 \\
    \end{bmatrix}=-J.
\end{equation*}

A transformation $H\in \A$ is called {\bf simple} if its matrix representation is
of the form
\begin{equation}\label{simple op}
    \begin{bmatrix}
      h & 0 \\
      0 & h \\
    \end{bmatrix}
    \qtext{where} h\in \A_0.
\end{equation}

We shall use a shorthand notation when dealing with an action by a simple transformation.
We shall write $h(f)$ or just $hf$ instead of
\begin{equation}
    \begin{bmatrix}\label{matrix h}
      h & 0 \\
      0 & h \\
    \end{bmatrix}
    \begin{bmatrix}
      f_1 \\
      f_2 \\
    \end{bmatrix}
    \fa
    \begin{bmatrix}
      f_1 \\
      f_2 \\
    \end{bmatrix}
    =f\in \E
\end{equation}
\begin{pro}[Commutativity of J with any simple transformation]
Every simple transformation $H\in \A$ commutes with transformation $J$
that is
\begin{equation*}
    H\circ J=J\circ H
\end{equation*}
In shorthand notation we can write $hJ=Jh$ if no confusion is possible.
\end{pro}

\bp
    The proof is obvious and we leave it to the reader.
\ep

Compare the above constructions with the arguments
presented in Kaiser \cite{kaiser1}, page 207.

\begin{thm}[Homogenous Maxwell equations imply homogenous wave equation]
Let $E$ and $H$ be two vector distributions from the space $\D'_3.$
If they satisfy the homogenous system of  Maxwell equations
\begin{equation*}
    \begin{split}
    &\nabla\times E=-DB,\quad \nabla\cdot E=0,\\
    &\nabla\times B=+DE,\quad \nabla\cdot B=0,\\
    \end{split}
\end{equation*}
then they satisfy the wave equations
\begin{equation*}
    \begin{split}
    &(\nabla^2-D^2)E=0,\\
    &(\nabla^2-D^2)B=0.\\
\end{split}
\end{equation*}
\end{thm}
\bigskip

\bp
Let
\begin{equation*}
    f=
    \begin{bmatrix}
      E \\
      B \\
    \end{bmatrix}
\end{equation*}
Then Maxwell's equations can be written as
\begin{equation*}
 (a)\qquad   (\nabla\times)f=DJf,\quad(b)\qquad (\nabla\cdot)f=0.
\end{equation*}
Applying operator $J^{-1}D=-JD$ to both sides of equation $(a)$ we get
\begin{equation}\label{D squared f}
    -JD(\nabla\times)f=D^2f.
\end{equation}
On the other hand applying the simple operator
$(\nabla\times)$ to both sides of equation $(a)$
and using the commutativity of the operators $J,D,(\nabla\times) $ we get
\begin{equation}\label{curl squared f}
    (\nabla\times)^2f=(\nabla\times)JDf=JD(\nabla\times)f.
\end{equation}
Since it is easy to verify that
\begin{equation}\label{formula for curl squared}
    (\nabla\times)^2f=\nabla((\nabla\cdot)f)-\nabla^2f\fa f\in \E,
\end{equation}
observing that the terms on the right side represent differential operators
generated by quadratic homogenous polynomials and the above identity holds
for any test function from $\D_3,$
by Maxwell's equation $(b)$ we get that the solution $f$ of the system
of Maxwell equations must satisfy the equation
\begin{equation*}
    (\nabla\times)^2f=-\nabla^2f.
\end{equation*}
Thus from equations \ref{D squared f},
\ref{curl squared f}, and \ref{formula for curl squared}
follows that the solution $f$ of the system of Maxwell equations must
satisfy the wave equation
\begin{equation*}
    \nabla^2f-D^2f=0
\end{equation*}
and this completes the proof.
\ep

\bigskip

\begin{pro}Assume that $E$ and $B$ are two, twice continuously differentiable,
vector fields from an open set $G$
in the space $R^4$ into $R^3$ satisfying the homogenous system of Maxwell's
equations
\begin{equation*}
    \begin{split}
    &\nabla\times E=-DB,\qquad \nabla\cdot E=0,\\
    &\nabla\times B=+DE,\qquad \nabla\cdot B=0,\\
    \end{split}
\end{equation*}
Then these fields satisfy the homogenous wave equations
\begin{equation*}
    \nabla^2E-D^2E=0,\qquad \nabla^2B-D^2B=0.
\end{equation*}
\end{pro}

\bigskip
\bp
    The proof follows from the previous theorem and from the
    fact that for continuous functions with continuous partials
    their partial derivatives in the sense of generalized functions
    coincide with the corresponding ordinary partial derivatives
    that is the weak derivatives coincides with the strong derivatives,
    see Theorem (\ref{weak and strong derivatives coincide}).
\ep


\section{Maxwell's mathematical experiment}
At the beginning of the 19-th century the magnetic and electric phenomena
were considered as independent. Ampere's law that electric current
can produce a magnetic field was the first breakthrough in
this field.

Around 1830 M. Faraday confirmed validity of Ampere's law.
Further he discovered that changing magnetic
flux can induce an electric current. He established the principle of
electric induction leading to transformers of alternating current.

In 1846 Faraday wrote a paper {\em Thoughts on ray vibrations.}
In his imagination he saw interaction of gravitational, electric, and magnetic
forces along the force lines,
which connected particles and masses together, contrary to prevailing Newtonian view
of actions at a distance. Faraday in his work suggested that light itself
is some kind of vibration in the lines of force. He postulated that
vibrating charge may produce vibrations in the lines of force.

J. C. Maxwell was inspired by Faraday's work and after augmenting Ampere's law
he discovered the specific velocity $c$
associated with electromagnetism that is now known as
the speed of light and proved that electromagnetic waves in free space
propagate with velocity $c.$

We can perform this experiment now in a more general setting
and a greater generality. Maxwell had done it
for harmonic waves in electromagnetic field.

Let us assume that we are working in some Lorentzian frame
and we have some physical quantity $u,$ it could be a component
of the electric or magnetic field, gravity field, or any other quantity,
that satisfies the homogenous wave equation in some open set $G\in R^4.$
We can select a rectangular subset of $G,$ so let us assume that
$G$ is in rectangular form, that is,
it is representable as a Cartesian product of 4 intervals.
The wave equation in standard g-m-s units for quantity $u$
will look as follows
\begin{equation}\label{wave velocity v}
    \nabla^2 u-\frac{1}{c^2}\frac{\partial^2}{\partial t^2}u=0\fa (r,t)\in G.
\end{equation}
We may assume without loss of generality that the origin of our
Lorentzian frame is in the center of the 4-dimensional cube $G.$
Take any function $f(t)$ of class $C^2$ on $R$ and consider a wave that
this function generates by means of the
formula
\begin{equation*}
    u(y_1,y_2,y_3,t)=f(y_1-vt)\fa (y_1,y_2,y_3,t)\in G.
\end{equation*}
At time $t=0$ the shape of the wave has graph represented by function the $t\mapsto f(t).$
To visualize this wave we may assume that
the wave looks like tsunami wave with the crest at $t=0.$
At any time $t$ the crest is at position $y_1=vt.$ Thus the wave moves with velocity
$v.$

We want to find when such a wave will satisfy our wave equation. Notice
that $u$ does not depend on $y_2$ and $y_3$ thus
\begin{equation*}
    \nabla^2 u=f''(y_1-vt)\fa (y_1,y_2,y_3,t)\in G.
\end{equation*}
and
\begin{equation*}
    \frac{\partial^2}{\partial t^2}u=v^2f''(y_1-vt)\fa (y_1,y_2,y_3,t)\in G.
\end{equation*}
The wave equation (\ref{wave velocity v}) will be satisfied for any
function of class $C^2$ if and only if $c^2=v^2$ that is $v=+c$ or $v=-c.$
Thus such waves propagate with velocity $c.$ This is the essence of
Maxwell's mathematical experiment. Hence we have the following corollary
\begin{cor}[Maxwell's experiment]
\label{Maxwell's experiment}
Assume that in a Lorentzian frame we have some physical quantity $u$
satisfying the homogenous wave equation (\ref{wave velocity v}), then waves
of that quantity propagate in free space with
velocity of light $c.$
\end{cor}

%
%

\bigskip
\section{Einstein's formulas for transformation\\ of electromagnetic field}
\bigskip


Lorentz \cite{lorentz1}
and Einstein \cite{einstein2a}, Part 2, section 6, established that
a pair of fields $$E=(E_1,E_2,E_3)\qtext{and}B=(B_1,B_2,B_3)$$
that satisfy homogenous system
of Maxwell equations in an open set $G$ transforms
as a part of an antisymmetric
tensor of second rank in Lorentzian space-time $R^3\times R.$ The matrix
of this tensor looks as follows
\begin{equation*}
    \begin{bmatrix}
      0 & +E_1 & +E_2 & +E_3 \\
      -E_1 & 0 & +B_3 & -B_2 \\
      -E_2 & -B_3 & 0 & +B_1 \\
      -E_3 & +B_2 & -B_1 & 0 \\
    \end{bmatrix}
\end{equation*}


For formulas for transformation of electromagnetic
field under Lorentzian change of coordinates see Feynman-Leighton-Sands
\cite{feyn2}, vol. 2, chapter 26. In particular formulas
for Lorentz transformations of the pair of fields $E$ and $B$
on page 26-9. Compare also with Dirac \cite{dirac}, section 23, page 42.

Following \cite{feyn2} assume that our orthogonal frames of reference are $S$ and $S'$
and the frame $S'$ moves along the first coordinate axis of $S$
with velocity $w=(w_1,0,0).$ Then the transformation of the components of the
vector  fields $$E=(E_1,E_2,E_3)\qtext{and}B=(B_1,B_2,B_3)$$
will look as follows
\begin{equation*}
    \label{table 26-3}\begin{tabular}{ll}
      $E'_1=E_1$ &\quad $B'_1=B_1$ \\
      $E'_2=\gamma{(E+w\times B)_2}$ &\quad $B'_2=\gamma{(B-w\times E)_2}$ \\
      $E'_3=\gamma{(E+w\times B)_3}$ &\quad $B'_3=\gamma{(B-w\times E)_3}$ \\
    \end{tabular}
\end{equation*}
where $\gamma=1/\sqrt{1-|w|^2}.$

Since $$w\times B=w\times (e\times E)=(w\cdot E)\,e-(w\cdot e)\,E$$
we get final formula involving just the vector fields $E$ and $e$
\begin{equation*}
 \begin{tabular}{l}
      $E'_1=E_1$  \\
      $E'_2=\gamma{(E_2+(w\cdot E)\,e_2-(w\cdot e)\,E_2)}
      =\gamma(E_2+w_1E_1\,e_2-w_1e_1\,E_2)$  \\
      $E'_3=\gamma{(E_3+(w\cdot E)\,e_3-(w\cdot e)\,E_3)}
      =\gamma(E_3+w_1E_1\,e_3-w_1e_1\,E_3)$  \\
\end{tabular}
\end{equation*}

\section{Bogdan-Feynman Theorem for a moving point mass}
\bigskip

From the theorems of the previous sections we can conclude that
the following theorem is true. Please notice that the physical
nature of the fields is completely irrelevant.

The important
fact is that we are working in a Lorentzian frame, in which
we are given an admissible trajectory, see definition (\ref{admissible trajectory}).
This trajectory alone generates in a unique way the system of the fundamental
fields by means of which we are able to construct fields that are preserved
by Lorentzian transformations.

As we have proved earlier
the notion of an admissible trajectory is invariant under any Lorentzian transformation
from one frame to any other frame moving with a constant velocity.
This fact is one of the consequences of
Einstein's formula on kinetic energy of a point mass.

We remind the reader that we are working in a fixed Lorentzian frame.
The position vector is denoted by $r_1\in R^3$ and time by $t\in R.$
The position axes are oriented so as to form right  screw orientation.
The units are selected so that the speed of light $c=1.$

Partial derivatives with respect to coordinates of $r_1$
are denoted by $D_1,\,D_2,\,D_3$ and with respect to time just by $D.$
The gradient differential operator is denoted by $\nabla=(D_1,D_2,D_3)$
and the D'Alembertian operator by $\Box^2=\nabla^2-D^2.$

\begin{thm}[Bogdan-Feynman Theorem]
\label{Bogdan-Feynman Theorem}
Assume that in a given Lorentzian frame the map $t\mapsto r_2(t)$
from $R$ to $R^3$ represents an
admissible trajectory of class $C^3.$ 
Assume that $G$ denotes the open set of points
that do not lie on the trajectory.
All the following field equations are satisfied on the entire set $G.$

Consider the pair of fields $E$ and $B$ over the set $G$
given by the formulas
\begin{equation*}
     E=u^2e+u^{-1}D(u^2e)+D^2e\qtext{and}B=e\times E
\end{equation*}
where $u$ and $e$ represent fundamental fields (\ref{Fundamental fields})
associated with the trajectory $r_2(t).$

Then this pair of fields will satisfy
the following homogenous system of Maxwell equations
\begin{equation*}
    \begin{split}
    &\nabla\times E=-DB,\quad \nabla\cdot E=0,\\
    &\nabla\times B=+DE,\quad \nabla\cdot B=0,\\
    \end{split}
\end{equation*}
and the homogenous wave equations
\begin{equation*}
    \Box^2 E=0,\qquad \Box^2 B=0.
\end{equation*}

Moreover Li\'{e}nard-Wiechert potentials, expressed in terms of the
fundamental fields as  $A=uzv$ and $\phi=uz,$ satisfy the homogenous
system of wave equations with Lorentz gauge formula
\begin{equation*}
    \Box^2 A=0,\quad \Box^2\phi=0,\quad \nabla\cdot A+D\phi=0
\end{equation*}
and generate the fields $E$ and $B$ by the formulas

\begin{equation*}
\begin{split}
    E&=-\nabla \phi-DA\qtext{and}B=\nabla\times A,\\
\end{split}
\end{equation*}

Finally we have the following explicit formula for the field $E$
in terms of the fundamental fields
\begin{equation*}\label{formula4F-b}
\begin{split}
    E&= -uz^2a   +uz^3 \langle e,a \rangle e -uz^3 \langle e,a \rangle v \\
    &\quad  +u^2z^3e-u^2z^3 \langle v,v \rangle e-u^2z^3v+u^2z^3 \langle v,v \rangle v.\\
\end{split}
\end{equation*}
\end{thm}
\bigskip

As a consequence of the above theorem the components of the quantities $E,$ $B,$
$A,$ and $\phi$ propagate in the Lorentzian frame with velocity of light $c=1.$
\bigskip

\section{Einstein's illusive gravity field}
\bigskip

Matter in the universe is distributed in chunks like atoms, particles, molecules,
or stellar objects, like planets, stars, galaxies, etc.
So it is natural to consider a system on $n$ bodies carrying charges
$q_j$ and having rest masses $m_{0j}.$

Einstein using general theory of relativity proved that waves in gravity
fields should also propagate in space with velocity $c$ like electromagnetic
waves. For reference see Einstein and Rosen \cite{einstein4}.
We shall assume that the field $E$ given by Newton-Feynman formula
has this property. We want to find out how to develop dynamics
in such fields.

Let us assume that we are working in a Lorentzian frame with units
selected so that the speed of light $c=1$ the electrostatic constant
in free space satisfies the condition $4\pi\epsilon_0=1$ and the unit
of mass is selected so that the gravitational constant $G=1.$

This means that if we keep meter as our unit of length the unit of time
is approximately 3.3 nanoseconds and the unit of mass is about $3.871$
metric tons and the unit of charge is $4\pi$ of coulombs.


Let us assume that we have a system consisting of $n$ bodies
indexed by $j,k=1,2,\ldots,n$ with rest masses $m_{0j}$ and charges $q_j.$
We want to find the
equations of motion for a time period  starting at time $t=s$ and ending at time $t=e.$
Assume that the bodies during this time period $[s,e]=I$
move without collision.

We shall follow here the constructions developed in Bogdan \cite{bogdan61}.
Now consider the trajectories.
Assume that we know the formulas for the trajectories
of the bodies up to time $t=s.$
We can setup the equations of motion of these bodies in their own force fields.
To this end let $r_{j}(t)$
be the trajectory of the $j$-th body.
Let $t_{jk}(t)$ be
the retarded time for a wave travelling at the speed of light to reach from
the $k$-th trajectory the $j$-th trajectory at the point $r_{j}(t)$ at time $t.$

This function must satisfy the equation
\begin{equation*}
 t-t_{jk}(t)=|r_{j}(t)-r_{k}(t_{jk}(t))|\fa t\le e.
\end{equation*}
As follows from Theorem (\ref{retarded time is unique})
on uniqueness of the retarded times,
the function $t_{jk}$  is uniquely determined by the above condition.

Put $T_{jk}=t-t_{jk}$ and call this function the delayed time.
Introduce a vector function $r_{jk}$ by the formula
\begin{equation*}
    r_{jk}(t)=r_j(t)-r_k(t_{jk}(t))\fa t\le e,\ j\neq k.
\end{equation*}
and the unit vector function $e_{jk}$ by
\begin{equation*}
    e_{jk}(t)=r_{jk}(t)/|r_{jk}(t)|\fa t\le e,\ j\neq k.
\end{equation*}

Since $T_{jk}=|r_j(t)-r_k(t_{jk}(t))|$
and by assumption the bodies do not collide for any
$t\in I$ the unit vector $e_{jk}(t)$ is well defined.
Let $v_{j}(t)$ be the velocity of the $j$-th body and let $y_{j}(t)$ denote
the state vector
$(r_{j}(t), v_{j}(t))$ and let $y(t)$ be the list of all state vectors $y_{j}(t)$
for $j=1,\ldots, n$.

Thus the vector function $y(t)$ has its values in the space $R^{6n}.$
By the norm of the state vector $y\in R^{6n}$
of the system $(r_j,v_j),$ where $j=1,\ldots, n,$ we shall understand
$$
    |y|=\max\set{|r_j|,|v_j|:\ j=1,\ldots, n},
$$
where $|r_j|$ and  $|v_j|$ denote the usual Euclidian norm in $R^3.$
\bigskip
Since the vector $r_{jk},$ and also the vector $e_{jk},$ is directed from
the $k$-th body towards the $j$-th body the Lorentz force induced by
the $k$-th body is given by the formula
\begin{equation*}
 q_j[E_{jk}+v_j\times(e_{jk}\times E_{jk})]
\end{equation*}
and Newton's force is given by
\begin{equation*}
 -m_{0k}E_{jk}
\end{equation*}
from the fact that force of action and force of reaction
are equal in magnitude but have opposite directions. Thus the total force $F_{jk}$
acting onto the $j$-th body by the $k$-th body is given by the formula
\begin{equation*}
 F_{jk}=q_j[E_{jk}+v_j\times(e_{jk}\times E_{jk})] -m_{0k}E_{jk}
\end{equation*}
\bigskip


Introduce an operator $H$ by the the formula
\begin{equation}\label{Opr H(jk)}
    H_{jk}(h)=q_{j}\{h+v_j\times(e_{jk}\times h)\}-m_{0k}h\fa h\in R^3.
\end{equation}
For fixed $y$ and $t$ and $j\neq k$  the operator $H_{jk}$ is linear with respect to
$h.$ Denote by $U$ the algebra of all linear transformations from the space $R^3$ into itself.

The total force acting onto the $j$-th body now can be represented in the form
\begin{equation}\label{total force on j}
    F_j=\sum^{k=n}_{k=1;k\neq j}H_{jk}(E_{jk})+F_{j\,0},
\end{equation}
where $F_{j\,0}$ represents external forces due to some external fields.
It could be a force field due to self action or due to loss of energy
through the electromagnetic radiation. We shall want to find conditions
that the fields should satisfy in order to have uniqueness of the
solution of the system of evolution equations.
\bigskip

Now for a body of rest mass $m_0$ moving under the influence of a force $F$
from Newton-Einstein formula, time derivative of the momentum is equal to
force, we must have
\begin{equation*}
    \dot{p}=F
\end{equation*}
where $\gamma=(1-|v|^2)^{-1/2}$ and $p=m_0\gamma v$ is the relativistic momentum.
Computing the derivative of $p$ with respect to time we get
\begin{equation*}
    \begin{split}
    \dot{p}&=m_0\gamma \dot{v}+m_0\dot{\gamma}v
    =m_0\gamma \dot{v}+m_0\gamma^3\dotp{v}{\dot{v}}v\\
    &
    =m_0\gamma ( \dot{v}+\gamma^2\dotp{v}{\dot{v}}v)=
    m( \dot{v}+\gamma^2\dotp{v}{\dot{v}}v)\\
    &=m\,\Gamma(v)(\dot{v}),\\
    \end{split}
\end{equation*}
where $m=m_0\gamma$ is the relativistic mass
and $\Gamma(v)$ is the linear transformation of $R^3$
into $R^3$ given by the formula
\begin{equation*}
 \Gamma(v)(h)=h+\gamma^2\dotp{v}{h}v\fa h\in R^3.
\end{equation*}
Notice that for fixed velocity $v$ the transformation $\Gamma(v)$
represents a symmetric, positive define transformation with eigenvalues
equal respectively to
\begin{equation*}
 (1+\gamma^2|v|^2),\ 1,\ 1.
\end{equation*}
Thus the inverse transformation $\hat{\Gamma}(v)$ exists and is also symmetric.
Since its eigenvalues are
\begin{equation*}
 (1+\gamma^2|v|^2)^{-1},\ 1,\ 1
\end{equation*}
and the norm of positive definite symmetric transformation is its maximal
eigenvalue, we must have
\begin{equation*}
 |\hat{\Gamma}(v)|=1\fa v\in R^3
\end{equation*}
and we have also
\begin{equation*}
 \hat{\Gamma}(v)\Gamma(v)=e\fa v\in R^3
\end{equation*}
where $e$ denotes here the identity transformation
in the algebra $Lin(R^3,R^3)$ that is $e(h)=h$
for all $h\in R^3.$

We shall call the function $v\mapsto \hat{\Gamma}$ the {\bf reciprocal to
the function} $\Gamma(v),$ since
\begin{equation*}
 \hat{\Gamma}(v)=[\Gamma(v)]^{-1}\fa v\in R^3.
\end{equation*}
\bigskip

Thus the Newton-Einstein equation for the $j$-th body can be written as
\begin{equation}\label{m Gamma=F}
    \Gamma(v_j)(\dot{v}_j)=\frac{1}{m_j }F_j\fa j=1,\ldots,n
\end{equation}

\bigskip
Since trajectories $y_j=(r_j,v_j)$ are admissible (\ref{admissible trajectory}),
for stopping time $t_1=e$ we must have
\begin{equation}\label{Lip condition for y(t)}
 \begin{split}
&\sup\set{|v_j(t)|:\ t\le e,\ j=1,2,\ldots,n}=q<1,\\
&\sup\set{|\dot{v}_j(t)|:\ t\le e,\ j=1,2,\ldots,n}=A<\infty\\
\end{split}
\end{equation}
\bigskip

Now put $x(t)=y(t)$ for $t\le s.$ Such trajectory will be called an initial
trajectory. Clearly $x$ is of class $C^1.$
We shall follow here the notation used in Bogdan \cite{bogdan61}.

\begin{defin}[Initial domain]
\label{Initial domain}
For a given initial trajectory $x$ and nonnegative numbers $q<1$ and $A<\infty$
and an interval $I=[s,e]$ define the set
\begin{equation*}
 D=D(x,q,A,I)
\end{equation*}
 of functions $y(t)=(r(t),v(t)),$ where $v(t)=\dot{r}(t),$
 extending the trajectory $x$ to the interval $I$
 and satisfying the inequalities
\begin{equation}
 \begin{split}
 |r_j(t)-r_j(\tilde{t})|&\le q|t-\tilde{t}|\fa t,\,\tilde{t}\le e.\\
 |v_j(t)-v_j(\tilde{t})|&\le A|t-\tilde{t}|\fa t,\,\tilde{t}\le e.\\
\end{split}
\end{equation}
Such a set $D$ will be called an {\bf initial domain} generated by the parameters
$x,q,A,I.$
\end{defin}

\begin{defin}[Space $C(I,Y)$ of continuous functions]
\label{space C(I,Y)}
For any interval $I$ and any Banach space $Y$ let $C(I,Y)$
denote the set of all continuous functions $f$ from $I$ into $Y.$
\end{defin}

In the case when the interval $I$ is closed and bounded the space $C(I,Y)$
with norm $\norm{f}=\sup\set{|f(t)|:\ t\in I}$ forms a Banach space.

The set $D(x,q,A,I)$
can be considered as a subset of the Banach space $C(I,R^{6n}).$
It is clear that the original trajectory $y$ of the entire system of bodies
belongs to the initial domain $D(x,q,A,I)$ for parameters $q$ and $A$ as
defined in (\ref{Lip condition for y(t)}).
\bigskip

\begin{defin}[Space $\mathcal{T}$ of admissible trajectories]
For fixed $n$ and an interval $I$ of the form $I=(-\infty,s\,]$
denote by $\mathcal{T}(I)$ the set of all admissible trajectories $x$ of the form
\begin{equation*}
 x_j(t)=(r_j(t),v_j(t))\qtext{ for }t\in I\text{ and }j=1,\ldots,n.
\end{equation*}
Trajectories $x$ of this form will be called {\bf admissible trajectories} on
interval $I$ and the set $\mathcal{T}(I)$ the {\bf space of admissible trajectories}
on $I.$

The union $\mathcal{T}$ of all such spaces $\mathcal{T}(I)$ will be
called the {\bf space of admissible trajectories}, that is
\begin{equation*}
 \mathcal{T}=\bigcup_I \mathcal{T}(I).
\end{equation*}
\end{defin}
\bigskip

Notice that every initial domain $D(x,q,A,[s,e])$ forms
a subset of the space $$\mathcal{T}((-\infty,e\,])\subset \mathcal{T}.$$

\begin{defin}[Space $\mathcal{U}(Y)$ of nonanticipating operators]
\label{Nonanticipating operator}
For fixed $n$ and a fixed Banach space $Y$
let $P$ denote an operator on the space $\mathcal{T}$ mapping
an admissible trajectory $x$ on an interval $I=(-\infty,d\,]$
into a continuous function from the space $C(I,Y).$

Such an operator is well defined if it has the following property:
for every two functions $y, z$ and every number
$d\in R$ the condition that $y(t)=z(t)$ for all $t\leq d,$ implies that the
images $P(y)(t)$ and $P(z)(t)$ coincide for all $t\leq d$.
Every such operator will be called a {\bf nonanticipating operator}
and the space $\mathcal{U}(Y)$ the {\bf space of nonanticipating operators}
with values in the space $Y.$
\end{defin}

\bigskip

The following theorem represents Th. 4.6 of Bogdan \cite{bogdan64}.

\begin{thm}[Initial domain is compact and convex]
Every initial domain $D(x,q,A,I)$ is compact
and convex in the Banach space $C=C(I,R^{6n}).$
\end{thm}
Thus every initial domain forms a closed set in the space $C(I,R^{6n}).$

\bigskip

\begin{defin}[Section $D_d\,$ of Initial Domain]
\label{Section $D_d$ of initial domain}
Let $D=D(x,q,A,I)$ be an initial domain with
$I=[s,e]$ For any $d\in [s,e]$ denote by $I_d$ the interval
$[s,e]$  and by $D_d$ the initial domain $D(x,q,A,I_d).$
Any such initial domain $D_d$ will be called a {\bf section of the initial domain} $D.$
\end{defin}

\bigskip

It is plain from the definition that every nonanticipating operator $P$ on an
initial domain $D$ induces a nonanticipating operator $P_d$ on any section
$D_d$ of the initial domain $D.$

\bigskip

\begin{defin}[Uniformly Lipschitzian Operator]
\label{Uniformly Lipschitzian Operator}
A nonanticipating operator $P$ from an
initial domain $D=D(x,q,A,I)$ into the space
$C(I,U),$ of continuous functions from the interval $I$ into a Banach
space $U,$ will be called {\bf uniformly Lipschitzian} if
there exists a constant $M$ such that for very $d\in I$ we have
\begin{equation*}
    \norm{P_d(y)-P_d(y^\sim)}_d\le M\norm{y-y^\sim}_d\fa y,\,y^\sim\in D_d,
\end{equation*}
where $\norm{\ }_d$ denotes the norm in the space $C(I_d,U).$
\end{defin}

\bigskip

Obviously the operators $y\into r_j$ and $y\into v_j$ are nonanticipating and uniformly
Lipschitzian on any initial domain $D.$
\bigskip

\begin{defin}[The space $Lip(D,U)$]
\label{The space $Lip(D,U)$}
Let $D=D(x,q,A,I)$ denote an initial domain and $U$ a Banach space. Let $Lip(D,U)$ denote
the space of all nonanticipating uniformly Lipschitzian operators
from the domain $D$ into the Banach space $C(I,U).$

If $P$ is an operator in $Lip(D,U),$ let
\begin{equation*}
    \begin{split}
    \norm{P}=&\sup\set{\norm{P(y)}:\,y\in D},\\
    \lip(P)=&\inf\set{M:\, \norm{P(y)-P(y^\sim)}_d
    \le M\norm{y-y^\sim}_d\fa y,y^\sim\in D,\ d\in I}.
    \end{split}
\end{equation*}
Clearly since $D$ is compact, this norm $\norm{P}$ is well defined.
The norm in the expression $\norm{P(y)}$ is
understood as the supremum norm in the space $C(I,U).$
Define
\begin{equation*}
    \norm{P}_\lip=\norm{P}+\lip(P)\fa P\in Lip(D,U).
\end{equation*}
The space $Lip(D,U)$ with norm $\norm{\ }_\lip$ will be called the
{\bf space of nonanticipating uniformly Lipschitzian operators.}
\end{defin}

\bigskip


\section{Nonanticipating differential equations}
\bigskip

Let $D=D(x,q,A,I)$ be an initial domain and $\Lambda\in Lip(D,R^{3n}).$
Consider the system of differential equations
\begin{equation}\label{equation with Lambda}
    \dot{v}_j(t)=\Lambda_j(y)(t)\fa  t\in I,\,j=1,\ldots,n.
\end{equation}
Introduce the operator $X$ who's component $X_j$ corresponding to the
$j$-th body is defined by the formulas
\begin{equation*}
     X_j(y)(t)=(v_j(t),\Lambda_j(y)(t)) \fa y=(r,v)\in D\,\text{ and } t\in I.
\end{equation*}
The right side of the equation (\ref{equation with Lambda})
represents a continuous function of the variable $t,$
so derivative on the left side also represents a contuous function.
Thus we can apply the integral to both sides to get an
equivalent equation
\begin{equation*}
    v_j(t)=v_j(s)+\int_s^t\Lambda_j(y)(u)\,du\fa t\in I,\,j=1,\ldots,n.
\end{equation*}
We also have
\begin{equation*}
    r_j(t)=r_j(s)+\int_s^tv_j(u)\,du\fa t\in I,\,j=1,\ldots,n.
\end{equation*}
Now if we define the operator $\Omega_j$ for all $y\in D$  by the formula
\begin{equation*}
    (r_j(s)+\int_s^tv_j(u)\,du,\ v_j(s)+\int_s^t\Lambda_j(y)(u)\,du)
\end{equation*}
we can rewrite the previous two equations in the form
\begin{equation*}
   y_j(t)= (r_j(t),v_j(t))=\Omega_j(y)(t)\fa t\in I,\,j=1,\ldots,n.
\end{equation*}
or even in a shorter form as
\begin{equation*}
    y=\Omega(y)
\end{equation*}
where $\Omega$ represents the list of operators $\Omega_j,\,(j=1,\ldots,n).$
This operator belongs to the space $Lip(D,R^{6n}).$
In this way we reduced the problem to a fixed point problem.

Now we are ready to prove that the fixed point problem has a
unique local solution. Compare the following theorem with Th. 7 of Bogdan \cite{bogdan61}.


\begin{thm}[Local existence and uniqueness]
\label{Local existence and uniqueness}
 Assume that the operator $\Lambda$ is nonanticipating and for every value of $A>0$
 there exists an interval $I=[s,e]$ such that $\Lambda$ restricted to the initial
 domain $D=D(x, q, A, I)$ yields an operator belonging to $Lip(D,R^{3n}).$

 Let $A_0>\abs{\Lambda(x)(s)}.$ Let $A>2A_0.$ Compactness of the initial
 domain $D=D(x,q,A,I)$ implies the compactness of its image $\Lambda(D).$
 Thus there exists a nondecreasing function $\omega(\delta),$ called modulus of continuity,
 such that $\omega(\delta)\into 0$
 when $\delta\into 0$ and
\begin{equation*}
 \abs{\Lambda(y)(t)-\Lambda(y)(t')}\le \omega(\abs{t-t'})\fa y\in D,\ t,t'\in I.
\end{equation*}
Let $\beta$ be such that $\omega(\beta)\le A_0$ and
$\lambda>\max\set{\lip(\Lambda_j):\,j=1,\ldots,n },$
where $\lip(\Lambda_j)$ denotes the Lipschitz constant
of the operator $\Lambda_j\in Lip(D,R^3).$

Let $J=[s,s+\delta]$ where $\delta$ is such that
\begin{equation*}
     0<\delta<\min\set{1,\beta,\lambda^{-1}}.
\end{equation*}
Then the transformation $\Omega$ maps the domain $D=D(x,q,A,J)$ into $D$
and forms a contraction map. Thus for the initial
trajectory $x$ of class $C^1$ there exists a unique local
solution in the domain $D$ to the evolution equation of the system of $n$ bodies
\begin{equation*}
    \dot{y}(t)=X(y)(t)\fa t\in I.
\end{equation*}
\end{thm}
\bp
    First let us establish that the operator $\Omega$ maps the
    initial domain $D$ into itself. Take any $y=(r,v)\in D$
    and let $\tilde{y}=(\tilde{r},\tilde{v})=\Omega(y).$
    For the position component we have
    \begin{equation*}
        \tilde{r}(t)=r(s)+\int_s^tv(u)\,du\fa t\le s+\delta.
    \end{equation*}
    and for the velocity component
    \begin{equation*}
        \tilde{v}(t)=v(s)+\int_s^t \ddot{r}(u)\,du\fa t\le s
    \end{equation*}
    and
    \begin{equation*}
        \tilde{v}(t)=v(s)+\int_s^t \Lambda(y)(u)\,du\fa s\le t\le s+\delta.
    \end{equation*}
    Notice that the above formulas represent a trajectory $\tilde{y}$  extending the
    initial trajectory $x.$

    To check the Lipschitz conditions notice that we have for the position component
    \begin{equation*}
        |\tilde{r}(t)-\tilde{r}(t^\sim)|=|\int_{t^{\sim}}^t v(u)\,du|
        \le  q |t-t^{\sim} |\fa t,t^{\sim}\in I.
    \end{equation*}
    For the velocity component we have
    \begin{equation*}
        \begin{split}
        \left|\tilde{v}(t)-\tilde{v}(t^\sim)\right|&=
        \left|\int_{t^{\sim}}^t\Lambda(y)(u)\,du\right|\\
        &\le\left|\int_{t^{\sim}}^t\Lambda(y)(s)\,du\right|+
            \left|\int_{t^{\sim}}^t\Lambda(y)(s)-\int_{t^{\sim}}^t\Lambda(y)(u)\,du\right|\\
        &\le\left|\int_{t^{\sim}}^t\Lambda(y)(s)\,du\right|+
            \left|\int_{t^{\sim}}^t(\Lambda(y)(s)-\Lambda(y)(u))\,du\right|\\
        &\le|\Lambda(x)(s)||t-t^{\sim} |+\omega(\delta)|t-t^{\sim} |\\
        &\le (A_0+\omega(\delta))|t-t^{\sim} |\le (A_0+\omega(\beta))|t-t^{\sim} |\\
        &\le 2A_0|t-t^{\sim} |\le A|t-t^{\sim} |\fa t,t^{\sim}\in I,
        \end{split}
    \end{equation*}
    Thus we have that $\tilde{y}\in D.$

\bigskip

    The transformation $\Omega$ is a contraction. Indeed, take any $y,y^\sim\in D$
    and consider their images $z=\Omega(y)$ and $z^\sim=\Omega(y^\sim).$
    Introducing the notation for components of $z=(r_1,v_1)$ and $z^\sim=(r_1^\sim,v_1^\sim)$
    we get
    \begin{equation*}
        \begin{split}
        |r_1(t)-r_1^\sim(t)|&=
        \left|\int_s^t(v(u)-v^\sim(u))\,du\right|=
        \left|\int_s^t\norm{v-v^\sim}\,du\right|\\
        &\le \delta\norm{y-y^\sim}\fa y,y^\sim \in D,\text{ and }t\in I
        \end{split}
    \end{equation*}
    and  for velocity component of $z,z^\sim$
    \begin{equation*}
        \begin{split}
        |v_1(t)-v_1^\sim(t)|&=
        \left|\int_s^t(\Lambda(y)(u)-\Lambda(y^\sim)(u))\,du\right|\\
        &\le \delta\lambda\norm{y-y^\sim}\fa y,y^\sim \in D,\text{ and }t\in I
        \end{split}
    \end{equation*}
    The above inequalities imply
    \begin{equation*}
        \norm{z-z^\sim}\le \max\set{\delta,\delta\lambda}\norm{y-y^\sim}\fa y,y^\sim\in D,
    \end{equation*}
    or equivalently
    \begin{equation*}
        \norm{\Omega(y)-\Omega(y^\sim)}\le q\norm{y-y^\sim}\fa y,y^\sim\in D,
    \end{equation*}
    where $q=\max\set{\delta,\delta\lambda}.$
    Since it follows from the assumptions of the theorem that the constant
    $q$ is less then $1,$ the transformation $\Omega$ represents a contraction mapping.

    Hence there exists a unique trajectory $y\in D$ such that $y=\Omega(y)$ or
    equivalently that
    \begin{equation*}
        \dot{y}(t)=X(y)(t)\fa t\in I.
    \end{equation*}
    This completes the proof of the theorem.
\ep

\bigskip


\section{Existence and Uniqueness of Global Solutions}

\bigskip
\begin{defin}[Operator with local uniqueness property]
\label{op with loc uniqueness}
Assume that we have a nonanticipating operator $X:\mathcal{T}\mapsto R^{6n}$ with the property
that for every admissible initial trajectory $x\in \mathcal{T}$
there exists a unique local solution
of the equation
\begin{equation*}
 \dot{y}(t)=X(y)(t) \fa t\in I
\end{equation*}
extending the trajectory $x.$
We shall say that such operator has the {\bf  local uniqueness property.}
\end{defin}
\bigskip

Assume that we found a solution $y$ of the differential equation in
an initial domain $D(x, q, A, I)$, where $I=[s,e].$
Taking the endpoint $e$ as our new starting point $s$ and the trajectory $y$ as
our new initial trajectory $x$ one can extend the solution onto a larger interval.
The object is to show that between all possible extensions there exists a
maximal extension. Moreover that the maximal extension is unique. Here we will make
the first step in this direction.

\bigskip

\begin{thm}[Uniqueness of extensions of solutions]
\label{Uniqueness of extensions of solutions}
Assume that a nonanticipating operator $X:\mathcal{T}\mapsto R^{6n}$
has a local uniqueness property for any starting time $s$
and any initial trajectory $x\in\mathcal{T}((-\infty,s\,]).$
Let $I$ be an interval starting at the point $s$.  On the right the interval may
be either open or closed, either bounded or unbounded.

If $y$  and $y^{\sim}$ are two
solutions of the differential equation
\begin{equation*}
 \dot{y}(t)= X(y)(t)\fa t\in I,
\end{equation*}
extending the initial trajectory $x$
onto the interval $I,$  then these trajectories are identical
on the entire interval $I$ that is
\begin{equation*}
    y(t)=y^{\sim}(t)\fa t\in I.
\end{equation*}
\end{thm}

\bigskip

\bp
    Let $I$ be an interval starting at the point $s$.
    On the right the interval may
    be either open or closed, either bounded or unbounded.

    Since difference of two continuous functions
    on the set $I$ yields  a continuous function, the function
    $\phi=y(t)-y^\sim(t)$ is
    continuous on $I.$
    Consider the set $J$ defined by
    \begin{equation*}
        J=\set{t\in I:\ \phi(u)=0\fa u\le t.}
    \end{equation*}
    Clearly $s\in J$ so the set $J$ is nonempty.

    Since $I$ is an interval and since the set $J$ with any two points $t_1,t_2\in J$
    such that $t_1<t_2$ contains all points $u$ such that $t_1<u<t_2,$ the set
    $J$ itself forms an interval.

    If the interval $J$ is, either unbounded on the right, or is bounded on the right but
    the least upper bound $\sup( J)$ coincides with
    the right end of the interval $I,$ we immediately see that $I=J.$

    So let us consider the case when $J$ is bounded on the right
    but $s_0=\sup( J)$ does not coincide with
    the right end of the interval $I.$ We shall prove that this leads to a
    contradiction.

    It follows from the continuity of the solutions $y_1$ and $y_2$
    that they coincide up to time $t=s_0$
    including $t=s_0$ and are of class $C^1.$ From our assumption about
    the point $s_0$ follows that it must lie inside of the interval $I.$
    Taking the part of the
    trajectories up to time $t=s_0$ as a new initial trajectory $x,$
    we can construct two unique local solutions
    defined on some intervals $I_1,\,I_2\subset I$ to the right of $s.$
    We may assume that $I_1=I_2,$
    otherwise we would take their intersection $I^\sim=I_1\cap I_2$ as a new interval.

    Let $\delta>0$ denote the length of the interval $I^\sim.$
    Since the solution of the equation
    is unique, we would have that
    \begin{equation*}
            y(t)=y^\sim(t)\fa t\in I_0=I\cap [s_0,s_0+\delta),
    \end{equation*}
    which would yield that $s_0+\delta\le s_0.$ A contradiction.
    Thus the interval $J$ must coincide with the entire interval $I.$
\ep

\bigskip

\begin{defin}[Maximal global solution]
\label{Maximal global solution}
Assume that $X$ is a nonanticipating operator with uniqueness property.
If between all possible solutions
extending an initial trajectory $x,$ there exists a solution with longest possible interval,
then such a solution is called {\it maximal global solution.}
\end{defin}

\bigskip

\begin{thm}[Existence and uniqueness of the maximal global solution]
\label{thm: global solution}
Assume that $X:\mathcal{T}\mapsto R^{6n}$ is
a nonanticipating operator with uniqueness property.

Then for every initial trajectory $x\in \mathcal{T},$
between all possible solutions $y$ of the differential equation
\begin{equation*}
    \dot{y}=X(y),
\end{equation*}
which extend $x,$ there exists a maximal global solution.
Moreover this solution is unique.
\end{thm}

\bigskip

\bp
    To prove the theorem let $\tilde{y}$ denote a trajectory,
    extending the initial trajectory $x,$ and
    defined on some interval $\tilde{I}=[\,s,\tilde{e}\,].$
    Assume that  $\tilde{y}$ represents a solution of the differential equation
    \begin{equation*}
    \dot{y}=X(y).
    \end{equation*}

    Take all such solutions with their intervals $\tilde{I}$ and let
    $I=\bigcup \tilde{I}$ be the union of all such intervals.
    If a point $t$ belongs to $I$ it belongs to some
    interval $\tilde{I}$ being the domain
    of a solution $\tilde{y}$. Put
    \begin{equation*}
        y(t)=\tilde{y}(t).
    \end{equation*}
    From the theorem on uniqueness of extensions follows that the value $y(t)$ is well
    defined, that is it does not depend on the choice of the function $\tilde{y}$.

    Indeed if $t$ belongs to any two intervals $\tilde{I}_1$ and $\tilde{I}_2$
    then on the intersection $J$
    of these intervals the corresponding solutions
    $\tilde{y}_1$ and $\tilde{y}_2$ must coincide
    in accord with the Theorem \ref{Uniqueness of extensions of solutions}.
    Thus we must have that $\tilde{y}_1(t)=\tilde{y}_2(t).$

    Moreover
    the obtained trajectory $y$ is the solution of our differential equation on the
    entire interval $I.$
    Since the graph of the function $y$ contains in it the graph of any
    other solution of our equation,
    the solution $y$ represents  a maximal global solution. Again from
    the Theorem \ref{Uniqueness of extensions of solutions} follows
    that this solution is unique.
\ep

\bigskip

\section{Spaces of nonanticipating uniformly Lipschitzian operators}

\bigskip

We shall remind the reader the definition \ref{The space $Lip(D,U)$}
of the space $Lip(D,U)$
of uniformly Lipschitzian operators on the initial domain $D.$
\bigskip

{\em Let $D=D(x,q,A,I)$ denote an initial domain and $U$ a Banach space.
Let $Lip(D,U)$ denote
the space of all nonanticipating uniformly Lipschitzian operators
from the domain $D$ into the Banach space $C(I,U),$ of all continuous functions from the
interval $I$ into the space $U.$
If $P$ is an operator in $Lip(D,U),$ let
\begin{equation*}
    \begin{split}
    \norm{P}=&\sup\set{\norm{P(y)}:\,y\in D},\\
    \lip(P)=&\inf\set{M:\, \norm{P(y)-P(y^\sim)}_d
    \le M\norm{y-y^\sim}_d\fa y,y^\sim\in D_d,\ d\in I}.
    \end{split}
\end{equation*}
Clearly, since $D$ is compact, this norm $\norm{P}$ is well defined.
The norm in the expression $\norm{P(y)}$ is
understood as the supremum norm in the space $C(I,U).$
Define
\begin{equation*}
    \norm{P}_\lip=\norm{P}+\lip(P)\fa P\in Lip(D,U).
\end{equation*}
The space $Lip(D,U)$ with norm $\norm{\ }_\lip$ is called the
{\bf space of nonanticipating uniformly Lipschitzian operators.}
}

\bigskip

\begin{pro}[The pair $(Lip(D,U),\norm{\ }_l)$ forms a Banach space]
\label{The pair Lip(D,U) with norm_l forms a Banach space}
\label{Lip(D,U) is Banach}
For every initial domain $D$ and for every Banach space $U$ the space $Lip(D,U),$
 of nonanticipating uniformly Lipschitzian operators, equipped with the norm
 $\norm{\ }_l$ forms a Banach space.
\end{pro}


\bp
    The proof follows from Th. 2.15 of Bogdan \cite{bogdan64}.
\ep

It follows from the above theorem that the space $Lip(D,U)$ is closed under addition and
scalar multiplication operations induced by respective operations in
the Banach space $U.$ In the sequel we will extend
this class of operations to include all Lipschitzian functions.
\bigskip

The following is Th. 6.10 of Bogdan \cite{bogdan64}.

\bigskip

\begin{thm}[Joint Continuity]
\label{Joint Continuity}
If $P\in Lip(D,U)$ is a nonanticipating uniformly Lipschitzian operator then the map
\begin{equation*}
    (t,y)\into P(y)(t)
\end{equation*}
is jointly continuous from the product $I\times D$ into the Banach space $U,$
that is for every point $(t_0,y_0)\in I\times D$ and for every $\varepsilon>0$
there exists $\delta>0$ such that
\begin{equation*}
    |P(y)(t)-P(y_0)(t_0)|\le \varepsilon\qtext{when} |t-t_0|\le \delta
    \text{ and }\norm{y-y_0}\le\delta\text{ and }(t,y)\in I\times D.
\end{equation*}
\end{thm}
\bigskip

\begin{cor}[Uniform continuity of $P(y)(t)$]
\label{Uniform continuity of $P(y)(t)$}
Assume that $P\in Lip(D,U).$
Since the set $I\times D,$ as product of compact sets, is itself compact.
Thus the function $(t,y)\into P(y)(t)$ is uniformly continuous on the set
$I\times D$ into the Banach space $U.$
\end{cor}

\bigskip

\begin{defin}[The set $ran(P)$ is the range of operator $P$]
\label{The set $ran(P)$ is the range of operator $P$}
Assume that $P\in Lip(D,U).$
For shorthand we shall write $ran(P)$ to denote the set
\begin{equation*}
    P(D)(I)=\set{P(y)(t):y\in D,\,t\in I}.
\end{equation*}
\end{defin}

\bigskip

\begin{cor}[Compactness of $ran(P)$] Assume that $P\in Lip(D,U).$
Since the set $I\times D$ is compact and
the function $(t,y)\into P(y)(t)$ is continuous on it, the
 set $ran(P),$ as an image of a compact
 set by means of a continuous map, forms a compact set in the Banach space $U.$
\end{cor}

\bigskip

\begin{defin}[Lipschitzian Operators over Cartesian Products]Let $$U_1,\ldots,U_n;\,Z$$
be some Banach spaces. We shall say that an operator $P,$ from a subset $G$ of the cartesian
product $U_1\times\ldots\times U_n$ into the Banach space $Z,$ is Lipschitzian on $G,$ if
for some constant $M$ we have
\begin{equation*}
    \norm{P(u_1,\ldots,u_n)- P(u^\sim_1,\ldots,u^\sim_n)}\le M
    (\norm{u_1-u^\sim_1}+\cdots+\norm{u_n-u^\sim_n})
\end{equation*}
for all $(u_1,\ldots,u_n),\,(u^\sim_1,\ldots,u^\sim_n)\in G.$
\end{defin}

\bigskip
The following is Th. 6.15  of Bogdan \cite{bogdan64}.

\begin{thm}[$Lip(D,U)$ spaces are closed under composition with Lipschitzian operators]
\label{Lip are closed under comp}
Let $D$ be an initial domain and $U_1,\ldots,U_n;\,Z$ be some Banach spaces.
Assume that $P_j\in Lip(D,U_j)$ for $j=1,\ldots,n$ and
$Q$ is a Lipschitzian operator from the product
$ran(P_1)\times\cdots\times ran(P_n)$ into the space $Z.$
Define the composed operator $P$ by the formula
\begin{equation*}
    P(y)(t)=Q(P_1(y)(t),\ldots,P_n(y)(t))\fa y\in D\text{ and }t\in I.
\end{equation*}
Then the operator $P$ belongs to the space $Lip(D,Z).$ We shall use
a shorthand notation for the operator $P=Q\circ(P_1,\ldots,P_n)$ or
shorter $P=Q(P_1,\ldots,P_n)$ when this
does not lead to a confusion.
\end{thm}

\bigskip

\begin{defin}[$n$-linear bounded operators ]Let $U_1,\ldots,U_n;\,Z$
be some Banach spaces. We shall say that an operator $P,$ from  the Cartesian
product $U_1\times\ldots\times U_n$ into the Banach space $Z,$ is {\bf n-linear} if
for every variable $u_j,$  where $j=1,\ldots,n,$ when other variables are fixed the map
\begin{equation*}
    u_j\into P(u_1,\ldots,u_j,\ldots,u_n)
\end{equation*}
is linear from the space $U_j$ into the space $Z.$
Such an operator is said to be {\bf bounded} if for some constant $M$ we have
\begin{equation*}
    |P(u_1,\ldots,u_n)|\le M\,|u_1|\,|u_2|,\ldots,|u_n|\fa u_1\in U_1,\ldots, u_n\in U_n.
\end{equation*}
\end{defin}

\bigskip

\begin{pro}[$n$-linear bounded operator is Lipschitzian on bounded sets]
If $P$ is an $n$-linear bounded operator from a Cartesian product $U_1\times\cdots\times U_n$
of Banach spaces into a Banach space $Z,$ then it is Lipschitzian on every set of the
form
\begin{equation*}
    \borel_\delta=\set{(u_1,\ldots,u_n)\in U_1\times\cdots\times U_n:\ |u_j|\le \delta,\ j=1,\ldots,n}
\end{equation*}
\end{pro}

\bigskip

\bp
    The proof is straightforward and we leave it to the reader.
\ep

\bigskip

\begin{cor}[Compositions with n-linear operators]
If $Q$ is an $n$-linear bounded operator from a Cartesian product $U_1\times\cdots\times U_n$
of Banach spaces into a Banach space $Z,$ and $P_j$ are in $Lip(D,U_j)$
 for $j=1,\ldots,n,$
then the composition $Q\circ(P_1,\ldots,P_n)$ belongs to the space $Lip(D,Z).$
\end{cor}

\bigskip

\bp
    Since the real-valued functions $u_j\into |u_j|$ are Lipschitzian on $U_j,$
    they are continuous. Thus on the compact set $ran(P_j)$ they attain their maximum
    $\delta_j.$ Put $\delta=\max\set{\delta_j:\ j=1,\ldots,n}.$

    Then the operator
    $Q$ is Lipschitzian on the bounded set $\borel_\delta$ and the operator
    representing the composition of operators
    \begin{equation*}
        P=Q\circ(P_1,\ldots,P_n)
    \end{equation*}
    is well defined and we have $P\in Lip(D,Z).$

\ep

\bigskip

\begin{defin}[Differentiable operators]
Let $f$ be an operator from an open set $G$ in a Banach space $U$ into a Banach space $Z.$
We shall say that the operator $f$ is differentiable at a point $a\in G$ if
there is a ball
\begin{equation*}
     \borel(a,r)=\set{u\in U:\ |u-a|< r} \subset G
\end{equation*}
 and a linear bounded operator $g\in Lin(U,Z)$ such that
\begin{equation*}
    |f(x)-f(a)-g(x-a)|\le o(|x-a|)\fa x\in\borel(a,r),
\end{equation*}
where $o(h)$ denotes some function of the variable $h\ge 0$ such that
\begin{equation*}
    o(h)/h\into 0\qtext{when}h\into 0.
\end{equation*}
Such an operator $g$ is unique, and is called the derivative of $f$ at the point $a,$
and it is denoted by $f'(a)=g.$
\end{defin}

\bigskip

It follows from the above definition that if the map $f$ has a derivative at a
point $a\in G$ then $f$ is continuous at $a.$
Notice also that the space $Lin(R,R)$ is isomorphic and isometric with the space
$R$ of reals.
\bigskip

The theorem, that follows, can be found in Cartan \cite[page 27, Th. 2.2.1]{cartan}.

\bigskip

\begin{thm}[Chain Rule]
Assume that $U,\,V,\,W$ are Banach spaces and $G\subset U$ and $H\subset V$ are open
sets. Let $f:G\into H$ and $g:H\into W$ be differentiable on their domains.
The composed function $h=f\circ g$ defined by
\begin{equation*}
    h(u)=f\circ g (u)=f(g(u))\fa u\in G
\end{equation*}
is differentiable and
\begin{equation*}
    h'(u)=f'(g(u))g'(u)\fa u\in G.
\end{equation*}
\end{thm}

\bigskip

The following theorem can be found in Cartan \cite[page 41, Th. 3.3.2]{cartan}.

\bigskip

\begin{thm}[Maps with bounded derivative on convex sets are Lipschitzian]
Let $f$ be an operator from an open set $G$ in a Banach space $U$ into a Banach space $Z.$
Assume that the derivative $f'(u)$ exists at every point $u\in G.$
Assume that for some convex set $W\subset G$ and a constant $M$ we have
\begin{equation*}
    |f'(u)|\le M\fa u\in W.
\end{equation*}
Then
\begin{equation*}
    |f(u)-f(u^\sim)|\le M\,|u-u^\sim|\fa u,u^\sim \in W.
\end{equation*}
\end{thm}

\bigskip

The space $Lin(U,U)$ of linear bounded operators from a Banach space $U$ into itself
beside being closed under addition and scalar multiplication is also closed under
the composition of operators: $P\circ Q\in Lin(U,U)$ for all $P,Q\in Lin(U,U).$
This operation has the property that $|P\circ Q|\le |P|\,|Q|.$

\bigskip

\begin{defin}[Banach Algebras]
A Banach space $U$ is called a {\bf Banach algebra,} if it is equipped with
a bilinear operation $(u,w)\into u\,w,$ from the product $U\times U$
into $U,$
that is associative: $(u\,w)\,z=u\,(w\,z),$
and such that
\begin{equation*}
    |u\,w|\le |u|\,|w|\fa u,\,w\in U.
\end{equation*}
If in addition there is an element $e$ in $U$ such that $e\,u=u\,e=u$ for all $u\in U,$
then such an algebra is called a {\bf Banach algebra with unit.} In such a case
an element $u$ is called invertible if for some $w\in U,$ called the inverse of $u,$ we have
\begin{equation*}
    u\,w=w\,u=e.
\end{equation*}
The unit element and the inverse elements are unique. We denote the inverse of $u$ by
$u^{-1}.$
\end{defin}

\bigskip

The following proposition is very useful in numerical computations of
the inverse transformations. It represents a simple application of
the Banach fixed point theorem and it is essential in the development of
the theory presented in this paper.
\bigskip

It represents Pro. 6.25 of Bogdan \cite{bogdan64}.
\bigskip
\begin{pro}[Inverse $(e-v)^{-1}$ exist if $|v|<1$]
\label{inverse (e-v)}
Let $U$ be a Banach algebra with unit. Then for every element $v\in U$ such that
$|v|<1$ the inverse $w=(e-v)^{-1}$ exists. It is a fixed point of the operator $f$
defined by the following formula
\begin{equation*}
    f(u)=e+vu\fa u\in U.
\end{equation*}
Moreover the sequence $w_n=f(w_{n-1})$ where $w_0=0$ is of the form
\begin{equation*}
    w_n=e+v+v^2+\cdots+v^n=\sum_{0\le k\le n}v^k\fa n>0
\end{equation*}
and we have the following estimate for the distance of the fixed point $w$ and
the approximation $w_n$
\begin{equation*}
    |w-w_n|\le \frac{|v|^n}{1-|v|}\fa n>0,
\end{equation*}
and we also have an explicit formula for the inverse element $w$ as the sum of an
absolutely convergent series
\begin{equation*}
    w=e+v+v^2+\cdots=\sum_{n\ge 0}v^n.
\end{equation*}
\end{pro}

\bigskip

Notice that in any Banach algebra $Lin(U,U)$ the identity map: $e(u)=u$ for all $u\in U,$
is the unit element.
It is easy to prove that the norm of the unit element $e$
in any Banach algebra is equal to 1, $|e|=1.$ Therefore for every invertible
element $u$ we have $1=|u\,u^{-1}|\le |u|\,|u^{-1}|,$ thus $ |u^{-1}|>0.$

\bigskip

The following is Pro. 6.26 of Bogdan \cite{bogdan64}.

\begin{pro}[When the inverse map $u\mapsto u^{-1}$ is Lipschitzian?]
\label{When the inverse map is Lipschitzian?}
Assume that $U$ is a Banach algebra with unit.
\begin{enumerate}
\item If for some $u_0$ the inverse $u_0^{-1}$
exists then every element in the ball $$\borel(u_0,r)=\set{u\in U:\,|u-u_0|<r },$$
where $r=1/|u_0^{-1}|,$ has an inverse.

\item The domain $G$ of existence of the inverse $u^{-1}$
is an open set and the function $f(u)=u^{-1}$ is continuous on $G.$

\item The function $f(u)=u^{-1}$ is differentiable on $G$ and
$$f'(u)(h)=-u^{-1}h\,u^{-1}\fa h\in U\text{ and }u\in G.$$

\item If $W$ is a compact convex set such that $W\subset G,$
then $f(u)$ is Lipschitzian on $W.$
\end{enumerate}
\end{pro}

\bigskip

\begin{thm}[If reciprocal $g$ of $f\in Lip(D,U)$ exists then $g\in Lip(D,U)$]
\label{reciprocal of f is Lip}
Let $D$ be any initial domain and $U$ a Banach algebra with unit $e.$
Then if $f$ is a uniformly Lipschitzian function from the domain $D$
into the algebra $U$ and
\begin{equation*}
 [g(y)(t)]\,[f(y)(t)]=e\fa y\in D\text{ and }t\in I
\end{equation*}
then $g\in Lip(D,U).$
\end{thm}
\bigskip

\bp
    By continuity of the functions $(y,t)\mapsto f(y)(t),$ and $x\mapsto x^{-1},$
    and the norm $y\mapsto \abs{y}$ and compactness of the set $D\times I$
    we get the existence of a constant $M$ such that
\begin{equation*}
     \abs{\,[f(y)(t)]^{-1}}\le M\fa y\in D\text{ and }t\in I.
\end{equation*}

    From the algebraic identity
\begin{equation*}
     x_1^{-1}-x_2^{-1}=x_1^{-1}(x_2-x_1)x_2^{-1}
\end{equation*}
    we get
\begin{equation*}
\begin{split}
     \abs{g(y_1)(t)-g(y_2)(t)}&=\left|[f(y_1)(t)]^{-1}
     [f(y_2)(t)-f(y_1)(t) ][f(y_2)(t)]^{-1}\right|\\
     &\le M^2 l(f)\norm{y_1-y_2}_d\fa y_1,y_2\in D_d\text{ and }t\in I_d,\,d\in I
\end{split}
\end{equation*}

Therefore
\begin{equation*}
     l(g)\le M^2 l(f)
\end{equation*}
and this implies that $g\in Lip(D,U).$
\ep
\bigskip

\begin{thm}[$f\in Lip(D,R^k)\Leftrightarrow f_j\in Lip(D,R)\ \forall\ j=1,\ldots,k$]
Let $f$ be a map from an initial domain $D$ into the space $R^k.$
Assume that $f=(f_1,\ldots,f_k)$ where $f_j$ is from $D$ into the reals $R$
for all $j=1,\ldots,k.$

Then $f$ belongs to the space $Lip(D,R^k)$ if and only if
each component $f_j$ belongs to the space $Lip(D,R).$
\end{thm}

\bigskip
\bp
    The proof is obvious and we leave  it to the reader.
\ep
\bigskip

\begin{thm}[$f\in C^1(R^k,R^m)\text{ and }g\in Lip(D,R^k)\Rightarrow f\circ g\in Lip(D,R^m)$]
Assume that $f$ is a continuous function from the space $R^k$ into the space $R^m$
having continuous partial derivatives of order 1.

Then if $g$ is a uniformly Lipschitzian function from an initial domain $D$ into
the space $R^k$ then the composed function $f\circ g$ is uniformly
Lipschitzian from $D$ into the space $R^m.$
\end{thm}
\bigskip

\bp
    By assumption the derivative $f'(x)$ exists and is continuous from $R^k$
    into the space $Lin(R^k,R^m)$ of linear transformations from $R^k$
    into the space $R^m.$ Thus on every closed bounded convex set
    the map $x\mapsto \abs{f'(x)}$ is bounded. Thus from Th. 3.3.2, page 41,
    of Cartan \cite{cartan} follows that $f$ is Lipschitzian on every such set.

    Therefore from Th. \ref{Lip are closed under comp} follows that
    $f\,\circ g$ is uniformly Lipschitzian on the initial domain $D.$
\ep
\bigskip

As a consequence of the above theorem we can conclude that for any polynomial $p$
the following is true.

\begin{cor}[$x_j\in Lip(D,R)\ \forall\ j \Rightarrow p(x_1,\ldots,x_k)\in Lip(D,R) $]
Assume that $D$ is an initial domain.
Assume that $p(x)$ is a polynomial of coordinates $x_j$ of the vector $x.$
If components $x_j=P_j\in Lip(D,R)$
then $p\circ (x_1,\ldots,x_k)$ is in the space $Lip(D,R).$
\end{cor}


\section{Initial Domains with Positive Separation}

\bigskip
In this section we will derive some estimates involving the speed $c$ of light,
so for a while we will explicitly use this constant.
Only in later sections it will be more convenient to change the units.
Compare the following developments with constructions in Bogdan \cite{bogdan61}
in particular with results in Th. 10 and Th. 11.

Let $D=D(x,q,A,I)$ be an initial domain.
In this section we will develop tools to study further the properties
of the spaces $Lip(D,U)$ that are need for analysis of operators
generated by moving bodies in gravitational and electromagnetic
fields.

\bigskip

\begin{defin}[Lattice Operations]
\label{Lattice Operations}
Assume that $R$ as before denotes the field of real numbers.
Introduce operations $\vee$ and $\wedge$ from $R\times R$ into $R$
by the formulas
\begin{equation*}
    p\vee r=\max\set{p,r}=\frac{1}{2}(p+r+|p-r|),\quad
    p\wedge r=\min\set{p,r}=\frac{1}{2}(p+r-|p-r|).
\end{equation*}
Any linear space $U$ of real-valued functions closed under composition
with these two operations will be called a {\bf linear lattice.}

Notice that if $f\in U$ then the absolute value $|f|$
of the the function $f$ is also in $U,$
since $|f|=(-f)\vee f.$ Now,
if in addition the space $U$ is a Banach space with a norm $\norm{\ },$
satisfying the following implication
\begin{equation*}
    g=|f| \impl \norm{g}\le \norm{f}\fa f\in U,
\end{equation*}
then $U$ is called a {Banach linear lattice.} It is easy to see that
the last condition guarantees that the lattice operations are continuous
in the norm topology.
\end{defin}

\bigskip

The operations $\vee$ and $\wedge$ are commutative, associative, and
Lipschitzian, since the unary operation $p\into |p|$ is Lipschitzian.
The following theorem is useful in getting estimates of Lipschitz
constant for operators in various spaces $Lip(D,U).$

\bigskip

The following represents Th. 7.2 of Bogdan \cite{bogdan64}.

\begin{thm}[$Lip(D,R)$ is a Banach linear lattice and a Banach algebra]
\label{Lip(D,R)is ll and B-algebra}
For every initial domain $D$ the Banach space $Lip(D,R),$ with the norm
defined by
\begin{equation*}
    \norm{P}_\lip=\norm{P}+\lip(P)\fa P\in Lip(D,R),
\end{equation*}
where
\begin{equation*}
    \begin{split}
    \norm{P}=&\sup\set{\norm{P(y)}:\,y\in D},\\
    \lip(P)=&\inf\set{M:\, \norm{P(y)-P(y^\sim)}_d
    \le M\norm{y-y^\sim}_d\fa y,y^\sim\in D_d,\,d\in I},
    \end{split}
\end{equation*}
forms a Banach linear lattice and a Banach algebra with unit.
\end{thm}


In the sequel we will need a separation  and minimal time delay operators.

\begin{defin}[Separation and Minimal Time Delay Operators]
\label{Sep and Mini Time Del}
Assume that $$D=D(x,q,A,I)$$ represents an initial domain with $I=[s,e].$
Define the operator $\mathrm{sep},$ called the
{\bf separation} operator,  by the formula
\[
\mathrm{sep}(y)(t)=
\min\{|r_{j}(t)-r_{k}(t)|:\, j, k=1,\ldots, n,j\neq k\}\fa y\in D,\,t\le e,
\]
and the operator $\mathrm{mtd},$ called the
{\bf minimal time delay} operator, by the formula
\[
\mathrm{mtd}(y)(t)=\min\{T_{jk}(t):\, j, k=1,\ldots, n,j\neq k\}\fa y\in D,\,t\le e.
\]
\end{defin}

\bigskip

The following represents Pro. 7.4 of Bogdan \cite{bogdan64}.

\begin{pro}[Operators $\ \mathrm{sep},\,\mathrm{mtd}\ $ belong to $Lip(D,R)$]
\label{Operators sep and mtd belong to Lip(D,R)}
The separation operators $\mathrm{sep}$ and $\mathrm{mtd},$ when the trajectories $y\in D$
are restricted to interval $I,$
belong to the space $Lip(D,R)$
of nonanticipating uniformly Lipschitzian operators.
\end{pro}

\bp
    The proof follows from theorem \ref{Lip(D,R)is ll and B-algebra}.
\ep

\begin{defin}[Time Delay Bound and Separation Bound]
Let $D=D(x,q,A,I)$ be an initial domain.
The {\bf time delay bound} $t_b$ for the domain $D$ is defined by
\[
    t_b=\inf\set{\mathrm{mtd}(y)(t):\ y\in D,\ t\in I}.
\]
Similarly we define the {\bf separation bound} $s_b$ for the domain $D$ by
\[
    s_b=\inf\set{\mathrm{sep}(y)(t):\ y\in D,\ t\in I}.
\]
\end{defin}

Notice since ranges of the operators $\mathrm{mtd}$ and $\mathrm{sep}$
are compact the constants $t_b$ and $s_b$ are well defined and are finite.

\bigskip

Notice that $t_b=0$ if and only if $s_b=0.$ Indeed since the
map $(y,t)\into \mathrm{mtd}(y)(t)$ is continuous and the set $D\times I$
is compact, there exist a trajectory $y\in D$ and a point $t\in I$
and a pair of indexes $j,k$ such that
for some $j,k$ we have
\begin{equation*}
    0=t_b=T_{jk}(t)=\frac{1}{c}|r_j(t)-r_k(t-T_{jk}(t))|
    =\frac{1}{c}|r_j(t)-r_k(t)|\ge\frac{1}{c}\mathrm{sep}(y)(t)\ge \frac{1}{c}s_b\ge 0,
\end{equation*}
thus $s_b=0.$ The converse that $s_b=0$ implies $t_b=0$ can be proved similarly.

\bigskip

\begin{defin}[Domains with Positive Separation]
\label{Domains with Positive Separation}
Any initial domain $D$ such that $s_b>0$ will be called a {\bf domain with
positive separation.}
\end{defin}

\bigskip

The following represents Pro. 7.7 of Bogdan \cite{bogdan64}.

\begin{pro}
If $D=D(x,q,A,I)$ is an initial domain with positive separation and $I=[s,e]$,
then for every $d$ such that $0<d-s<t_b$ and $d\le e$ the emission time operators $t_{jk}$
satisfy the inequality
\begin{equation*}
    t_{jk}(t)=t-T_{jk}(t)\le s\fa y\in D_d\text{ and } t\in [s,d].
\end{equation*}
\end{pro}

\bigskip

\bp
    Notice that for the section $D_d$ of the domain $D$
    the time interval is $[s,d].$
    Thus for any $t\in[s,d]$ and any trajectory $y\in D_d$ we have
    \begin{equation*}
        t_{jk}(t)=t-T_{jk}(t)\le d-t_b\le d-(d-s)= s.
    \end{equation*}
    This completes the proof.
\ep

\bigskip

The operators $y\into r_k(t_{jk}),$ and  $y\into v_k(t_{jk}),$ and $y\into \dot{v}_k(t_{jk}),$
are constant on the domain $D_d$ and they are uniquely determined
by the initial trajectory $x,$ since $t_{jk}(t)\le s$ and
by definition the initial trajectory $x$ has a continuous derivative. Therefore each, of the
above operators, represents a continuous function of the variable $t.$
Thus all the above operators belong to the space $Lip(D_d,R^3).$

Notice that the time delay bound $t_b$ for
the domain $D$ and the time delay bound $t'_b$ for its section $D_d$
satisfy the inequality $t_b\le t'_b.$
Thus taking eventually the section $D_d$ as our new domain we observe that
the following corollary must be true.

\bigskip

\begin{cor}[Operators $r_k(t_{jk}),\ v_k(t_{jk}),\ v'_k(t_{jk}),\ r_{jk}$ are in $Lip(D,R^3)$]
\label{Remote are in Lip(D,R^3)}
Assume that $D=D(x,q,A,I)$ is an initial domain with positive separation,
and with $I=[s,e].$
If we have that $0<e-s<t_b,$ then each of the following operators
\begin{equation*}
    r_k(t_{jk}),\ v_k(t_{jk}),\  v'_k(t_{jk}),\ r_{jk}=(r_j-r_k(t_{jk})),\ (v_j-v_k(t_{jk})),
\end{equation*}
where $j,k=1,\ldots,n,j\neq k,$
belongs to the space $Lip(D,R^3).$
\end{cor}

\bigskip

The following represents Th. 7.9 of Bogdan \cite{bogdan64}.

\begin{thm}[$s_b\le c(1+q)t_b$]
\label{S_b le to c(1+q)T_b}
Assume that $D=D(x, q, A, I)$ is an initial domain, and $s_b$ is
the separation bound for the domain $D,$ and $t_b$ is
the time delay bound for the domain $D.$
Then
\begin{equation*}
    s_b\le c(1+q)t_b.
\end{equation*}
\end{thm}

\bigskip

The following represents Th. 7.10 of Bogdan \cite{bogdan64}.

\begin{thm}[Lower estimate of separation bound $s_b$]
\label{Lower estimate of separation bound $s_b$}
Let $D=D(x, q, A, I)$  be an initial domain with $ I=[s, e]$ and
with positive initial separation $\mathrm{sep}(x)(s)>0.$
 If the length of the interval $I=e-s$ is such that
 \begin{equation*}
        e-s\le\frac{\mathrm{sep}(x)(s)}{3cq},
 \end{equation*}
 then  we have the following lower estimate for the separation bound $s_b$
 \begin{equation*}\label{lower estimate of s_b}
        \frac{\mathrm{sep}(x)(s)}{3}\le s_b.
 \end{equation*}
\end{thm}

\bigskip
The following represents Th. 7.11 of Bogdan \cite{bogdan64}.

\begin{thm}[Negative powers of time delays are in $Lip(D,R)$]
\label{Negative powers of T are in Lip}
Let $D=D(x, q, A, I)$  be an initial domain with positive separation.
Then for every pair of indexes $j\neq k$ the operator $y\into T^{-n}_{jk},$ where $n>0,$
is in $Lip(D,R).$ Its Lipschitz constant can be estimated by
\begin{equation}\label{negative powers of T}
    \lip\,(T^{-n}_{jk})\le nt_b^{-n-1}\lip\,(T_{jk})\le  \frac{2n}{c(1-q)t_b^{n+1}}.
\end{equation}
\end{thm}
\bigskip

The following represents Cor. 7.12 of Bogdan \cite{bogdan64}.

\begin{cor}[Operators $|r_{jk}|,\ |r_{jk}|^{-n},\ e_{jk}$]
\label{Operators |r_jk|, e_jk}
Assume that the initial domain $D=D(x,q,A,I)$ is with positive separation.
If the length of the interval $I=[ s,e]$ is such that $e-s<t_b,$
then  the operators
\begin{equation*}
    y\into |r_{jk}|,\quad   y\into |r_{jk}|^{-n},(n>0)\fa j\neq k
\end{equation*}
are in $Lip(D,R),$ and the unit operator
\begin{equation*}
    y\into e_{jk}=|r_{jk}|^{-1}r_{jk}\fa j\neq k
\end{equation*}
is in $Lip(D,R^3).$
\end{cor}




\section{Lip(D,U) spaces over regular initial domains }
\bigskip

\begin{defin}[Regular initial domain]
\label{Regular initial domain}
An initial domain $D=D(x,q,A,I),$ where $I=[s,e],$
with positive initial separation $\mathrm{sep}(x)(s)>0$ and such that
\begin{equation*}
    0 < e-s<\frac{\mathrm{sep}(x)(s)}{3(1+q)c}
\end{equation*}
will be called a {\bf regular initial domain.}
\end{defin}

\bigskip

Since on every regular initial domain $D$ we have the inequalities for the time delay bound
\begin{equation*}
    \frac{\mathrm{sep}(x)(s)}{3(1+q)c}\le t_b=\inf\set{\mathrm{tdb}(y)(t):\ y\in D,\,t\in I }
\end{equation*}
and for the separation bound
\begin{equation*}
    \frac{\mathrm{sep}(x)(s)}{3}\le s_b=\inf\set{\mathrm{sep}(y)(t):\ y\in D,\,t\in I },
\end{equation*}
the bounds $t_b$ and $s_b$ are positive. It will be convenient from now on to assume,
if not specified otherwise, that the domain $D$ is regular.

\bigskip
From now on we will assume that $c=1.$
We have already proved various parts of the following theorem but to be sure
that we did not overlooked any particular fundamental field we collet them
all into one theorem and go over the proofs again now that we have more tools to
do that.

\begin{thm}[Fundamental fields due to $k$-th body acting onto $j$-th body are\\
in $Lip(D,R^m)$ for $m=1$ or $m=3$]
Assume that $D=D(x,q,A,I)$ is a regular initial domain. Consider the fields
acting onto the $j$-th body by the $k$-th body corresponding to the
fundamental fields of the system of the trajectories.

Then the scalar fields
\begin{equation*}
 y\mapsto T_{jk},\ u_{jk},\ t_{jk},\ z_{jk}
\end{equation*}
all belong to the space $Lip(D,R)$ and the vector fields
\begin{equation*}
 y\mapsto r_j,\  v_j,\  r_{jk}=r_j-r_k(t_{jk}),\  e_{jk}=u_{jk}r_{jk},\  r_j(t_{jk}),\
v_j(t_{jk}),\ a_j(t_{jk})
\end{equation*}
all belong to the space $Lip(D,R^3).$
\end{thm}

\bp
    It is obvious that operators $ y\mapsto r_j$ and $ y\mapsto v_j$ are in $Lip(D,R^3).$
    The remote operators, as follows from (\ref{Remote are in Lip(D,R^3)}),
    \begin{equation*}
       y\mapsto r_j(t_{jk}),\ v_j(t_{jk}),\ a_j(t_{jk})
    \end{equation*}
    are also in $Lip(D,R^3).$
    From linearity of $Lip(D,R^3)$ follows that operator
    \begin{equation*}
       y\mapsto  r_{jk}=r_j-r_k(t_{jk})
    \end{equation*}
    is in $Lip(D,R^3).$
    The operator $y\mapsto T=\abs{r_j-r_k(t_{jk})}$ as a composition of
    Lipschitzian function $r\mapsto \abs{r}$ with $r_{jk}$ is in $Lip(D,R).$

    Since on a regular domain we have the estimate
    \begin{equation*}
        T_{jk}(y)(t)\ge t_b>0\fa y\in D\text{ and }t\in I
    \end{equation*}
    the operator $u_{jk}$ as the reciprocal operator to $T_{jk}$ is in $Lip(D,R).$
    Thus the operator $e_{jk}=u_{jk}r_{jk}$ considered as composition
    of operators $u_{jk}\in Lip(D,R)$ and operator $r_{jk}\in Lip(D,R^3)$
    with bilinear bounded operator $(\lambda,r)\mapsto \lambda\, r$ is in $Lip(D,R^3).$

    Similarly we can prove that operators $t_{jk}=t-T_{jk}$
    and $z_{jk}=(1-\dotp{e_{jk}}{v_{jk}})^{-1}$ are in $Lip(D,R).$
\ep
\bigskip

\begin{thm}[Newton-Feynman fields $E_{jk}$ are in $Lip(D,R^3)$ ]
Assume that $D=D(x,q,A,I)$ is a regular initial domain. Consider
the Newton-Feynman field $E_{jk}$
acting onto the $j$-th body by the $k$-th body of the system
$y\in D$ of the trajectories.
Then for every pair of indexes $j\neq k$ the field $E_{jk}\in Lip(D,R^3).$
\end{thm}
\bigskip

\bp
    According to Bogdan-Feynman Theorem (\ref{Bogdan-Feynman Theorem})
    the Newton-Feynman field $E$ generated by a single trajectory and expressed
    in terms of the fundamental fields is
    given by the formula
    \begin{equation}\label{N-F single field}
    \begin{split}
        E&= -uz^2a   +uz^3 \langle e,a \rangle e -uz^3 \langle e,a \rangle v \\
        &\quad  +u^2z^3e-u^2z^3 \langle v,v \rangle e-u^2z^3v+u^2z^3 \langle v,v \rangle v.\\
    \end{split}
    \end{equation}
    Thus the components of $E=(E_1,E_2,E_3)$
    are representable as polynomials of the components of the fundamental fields.
    Hence $E_{jk}$ is representable as a polynomial of the coordinates
    of all the fields appearing in the previous theorem. So each coordinate
    of the field $E_{jk}$ is in $Lip(D,R)$ so the field itself $E_{jk}$
    is in $Lip(D,R^3).$
\ep

\bigskip

\section{Existence of solutions to the evolution equations in force fields of
special theory of relativity}
\bigskip

We remind the reader that the Newton-Einstein (\ref{m Gamma=F})
equation for the $j$-th body can be written as
\begin{equation*}
    \Gamma(v_j)(\dot{v}_j)=\frac{1}{m_j }F_j\fa j=1,\ldots,n
\end{equation*}
or equivalently as
\begin{equation}\label{Gamma=F/m}
    \dot{v}_j=\frac{1}{m_j }\hat{\Gamma}(v_j)(F_j)\fa j=1,\ldots,n
\end{equation}

\begin{defin}[Regular nonanticipating operator]
By a {\bf regular nonanticipating} operator we shall
understand an operator $F:\T\mapsto R^{3n}$ such that for every regular
initial domain $D=D(x,q,A,I)$  the restriction of $F$ to $D$
represents a uniformly Lipschitzian operator, that is, it represents
an element of the space $Lip(D,R^{3n}).$
\end{defin}
\bigskip

Notice that in the following theorem the nature of the force operator
$F$ is not important. Important is just the mathematical assumption
that it represents a regular nonanticipating operator. This is
the property we were looking for.
\bigskip

\begin{thm}[Maximal solutions of equations of special theory of relativity]
Assume that the operator $F:\T\mapsto R^{3n}$ representing the force
field forms a regular nonanticipating operator.

Then for every admissible initial trajectory $x$ of the system of $n$ bodies
such that at the initial time $t=s$ the separation operator satisfies
the condition $$\mathrm{sep}(x)(s)>0,$$
there exists a unique maximal solution
$\ y\ $ to the Newton-Einstein system of equations 
\begin{equation*}
     \dot{v}_j=\frac{1}{m_j }\hat{\Gamma}(v_j)(F_j)\fa j=1,\ldots,n
\end{equation*}
satisfying the condition
\begin{equation*}
 \mathrm{sep}(y)(t)>0\fa t\in I
\end{equation*}
and extending the initial trajectory $x$ onto the time interval $I.$
\end{thm}

\bigskip

\bp
    Let $U$ denote the algebra of all linear transformations from $R^3$
    into $R^3.$ The map
\begin{equation*}
     g(r,u,x)= r\,u(x)\fa r\in R,\ u\in U,\ x\in R^3
\end{equation*}
is trilinear and bounded.

Take any regular initial domain $D=D(x,q,A,I).$
Consider first the map
$$y\mapsto m_j^{-1}=m_{0j}^{-1}\sqrt{(1-|v_j|^2)}.$$
Notice that $p_j(y)=(1-|v_j|^2)$ represents a polynomial of the coordinates of
the vector operator $v_j\in Lip(D,R^3).$ Thus $p_j\in Lip(D,R).$
We also have the estimate $a=(1-q^2)\le(1-|v_j|^2)\le 1.$
Since the square root function is differentiable on the closed interval $[a,1]$
and has continuous derivative, it is Lipschitzian. Thus
the operator $y\mapsto m_j^{-1}$ is in $Lip(D,R)$ for every $D.$


Consider the operator $y\into \Gamma(v_j).$ By definition
\begin{equation*}
 \Gamma(v_j)(h)=h+\gamma_j^2\dotp{v_j}{h}v_j\fa h\in R^3.
\end{equation*}
Notice $\gamma_j^2=p_j^{-1}.$ Since we established before that $p\in Lip(D,R)$
its reciprocal $\gamma^2$ is in $Lip(D,R).$
\bigskip

Now consider the 4-linear bounded map
\begin{equation*}
 S(r,v_1,v_2,h)=r\dotp{v_1}{h}v_2\fa r\in R;\ v_1,v_2,h\in R^3
\end{equation*}
and define map
\begin{equation*}
 \tilde{S}(r,v_1,v_2)(h)=r\dotp{v_1}{h}v_2\fa r\in R;\ v_1,v_2,h\in R^3
\end{equation*}
Clearly the map $\tilde{S}$ is trilinear and bounded and maps
the triple $(r,v_1,v_2)$ into an element of the algebra $U.$
Thus we have $ \tilde{S}(\gamma_j^2,v_j,v_j)\in Lip(D,U).$
Since the unit element $e$ of the algebra $U$ as a constant
operator belongs to $Lip(D,U)$ and
\begin{equation*}
 \Gamma(v_j)=e+\tilde{S}(\gamma_j^2,v_j,v_j)
\end{equation*}
we must have $\Gamma(v_j)\in Lip(D,U).$
Therefore the reciprocal operator $y\mapsto \hat{\Gamma}(v_j)$
must belong to $Lip(D,U)$ for every
regular initial domain $D.$

Hence the operator $\frac{1}{m_j }\hat{\Gamma}(v_j)(F_j)$ belongs to
the space $Lip(D,R^3)$
for every regular initial domain $D.$
Therefore the operator $y\mapsto\frac{1}{m_j }\hat{\Gamma}(v_j)(F_j)$
forms a regular nonanticipating operator.

The above on the basis of the theorem on global maximal
solution leads to the conclusion stated in the theorem:
There is a unique maximal solution satisfying the separation
condition and extending the initial trajectory.
\ep

\bigskip
\section{Existence of solutions to equations of gravitational electrodynamics
with external force fields}
\bigskip

We have introduced in (\ref{Opr H(jk)}) operators $H_{jk}$ by the the formula
\begin{equation}
    H_{jk}(h)=q_{j}\{h+v_j\times(e_{jk}\times h)\}-m_{0k}h\fa h\in R^3.
\end{equation}
These operators have the following property.

\begin{pro}[Operators $H_{jk}\in Lip(D,U)\ \forall\ j\neq k$]
The operators $H_{jk}$ for every pair of indexes $j\neq k$ are well defined
on any regular initial domain $D=D(x,q,A,I)$ and belong
to the space $Lip(D,U)$ where $U$ denotes the algebra of
all linear operators from $R^3$ into $R^3.$
Thus the operators $H_{jk}$ are regular nonanticipating operators.
\end{pro}

\bp
    The proof is similar to the proof concerning the operator $\hat{\Gamma}(v_j)$
    and we leave it to the reader.
\ep

\begin{defin}[Lorentz-Newton force operator]
By the Lorentz-Newton force operator we shall understand
the system of the operators, for j=1,\ldots,n, given by the formulas
\begin{equation}\label{total force on j-b}
    F_j=\sum^{k=n}_{k=1;k\neq j}H_{jk}(E_{jk})\fa j=1,\ldots,n,
\end{equation}
where
\begin{equation}\label{Opr H(jk)-b}
    H_{jk}(h)=q_{j}\{h+v_j\times(e_{jk}\times h)\}-m_{0k}h\fa h\in R^3
\end{equation}
and
$E_{jk}$ represents the Newton-Feynman field acting onto the $j$-th body
from the trajectory of the $k$-th body of the system.
\end{defin}

\begin{thm}[Equations of gravitational electrodynamics
with external force field]
Assume that $F_{j\,0}:\T\mapsto R^{3n}$ are regular nonanticipating
operators representing external force fields.
Assume that $F_j\,(j=1,\ldots,n)$ denotes the Lorentz-Newton force
operator.

Then for every admissible initial trajectory $x,$
defined on the time interval $(-\infty,s],$
with positive separation $\mathrm{sep}(x)(s)$ at time $t=s,$
there exist an interval $I$ staring at $s,$ such that
the Newton-Einstein system of differential equations
\begin{equation*}\label{Newton-E-L}
 \dot{p_j}(t)=F_j(t)+F_{j\,0}(t)\fa t\in I\text{ and }j=1,\ldots,n
\end{equation*}
has a unique maximal solution having positive separation
\begin{equation*}
 \mathrm{sep}(y)(t)>0\fa t\in I.
\end{equation*}
\end{thm}

\bp
    Let $U$ denote the algebra of all linear operators from $R^3$ into $R^3.$
    Notice that Lorentz-Newton force operator forms a regular nonanticipating operator.
    Indeed, for fixed $j\neq k$ the operator $H_{jk}\in Lip(D,U),$
    where $D=D(x,q,A,I).$

    The operator $E_{jk}\in Lip(D,R^3).$ Since the map $S(u,r)=u(r)$
    considered on the product $U\times R^3$ into $R^3$ is bilinear and bounded
    we have that $H_{jk}(E_{jk})\in Lip(D,R^3).$

    Now from linearity of the space $Lip(D,R^3)$ follows that the
    Lorentz-Newton force $F_j\in Lip(D,R^3).$
    Since the force $F_{j\,0}$ represents a
    regular nonanticipating operator the sum
    $\Lambda_j=F_{j}+F_{j\,0}$  belongs to $Lip(D,R^3).$
    Thus the equation (\ref{Newton-E-L}) is equivalent to
    equation
\begin{equation*}
     \dot{v}_j(t)=[\frac{1}{m_j}\Lambda_j](t)\fa t\in I\text{ and }j=1,\ldots,n.
\end{equation*}
    We have proved in the preceding section that such system of equations
    has a unique maximal solution satisfying the condition
    that separation between the bodies is positive at any time $t\in I.$
\ep


\section{Computer programs for DERIVE software system}
\bigskip

The results presented in this paper were
obtained with the help of an interactive, symbolic, DOS based, computer program
DERIVE, version 2.06, developed by Soft Warehouse Inc. in Honolulu, Hawaii.

Since the differential formulas obtained in this paper essentially
yield polynomials of the involved field variables, any
program that can handle operations on polynomials of an arbitrary number
of variables can be used.

For a reader who would be interested in verifying the formulas
we enclose the header and the actual programs used in the computations.

To understand these programs we need to introduce rudimentary
elements of grammar used by DERIVE.

\verb"<variable>" consists of a string of lower case letters a-z, digits 0-9,
and underscore character \verb"_." A variable must start with a letter.

\verb"<algebraic expression>" consists of variables joint by usual operations:
addition  \verb"+,"  subtraction  \verb"-,"  multiplication  \verb"*,"  division  \verb"/,"
and exponentiation  \verb"^."
Parenthesis  \verb"(  )"  can be used to indicate the order of operations.

 \verb"<assignment statement>"  has the form

 \verb"<variable>:=<algebraic expression><cr>"

where  \verb"<cr>"  denotes the carriage return character.

 \verb"<comment>"  consists of any text enclosed into double quotes \verb'"<text>"'.

Using tilde \verb"~" as the continuation character one can break
long assignment statements or comments between lines. On output DERIVE
breaks line after every 80 characters.

DERIVE accepts input files containing programs. For illustration assume
that we prepared using a plain ASCII text editor a file {\bf hh.mth}
containing the text as in the program below representing the header with
basic formulas. Notice that the sequence in which the formulas are
entered in the program is important. One should not use things
that where not defined in previous formulas.

To make a test to see that DERIVE understands the program
run DERIVE. On entering derive a window appears with menu commands
at the bottom of the window. Notice that each command has one
letter in upper case. To select such a command just press the
corresponding letter on the keyboard.

To load the program select from consecutive menus: Transfer,
Load, Derive, hh.
Press \verb"<cr>" that is \verb"<Enter>" on most keyboards.
DERIVE will load the file {\bf hh.mth} and the last statement
in the file will be highlighted. To browse through the statements
use directional keys \verb"<up>" and \verb"<down>" on the keyboard.
The selected statement will be highlighted. To enter a statement to
examine its components use directional keys \verb"<right>" and \verb"<left>"
and to enter or exit a subcomponent use the keys \verb"<up>" and \verb"<down>".

On a highlighted component one can perform any operation
from the menu. We need just one operation: Expand.
It permits one to expand a compound component into a polynomial
of variables that are involved in the component.

Expanded components are appended at the bottom of the file {\bf hh.mth}
To save the computations select from the consecutive menus commands:
Transfer, Save, Derive, \verb"<cr>". The file {\bf hh.mth} will be
saved to its original location overwriting the previous version.
To exit DERIVE select from the menu command Quit.

Using your favorite ASCII editor update the file by new
statements and comments. One can repeat this process until the goal
of the computations is met.

In the following programs the names of the variables were preserved,
if possible, as they appear in the paper with proper adjustments
to lower case letters.

To represent the dot product $\dotp{e}{v}$ of vectors
 $e$ and $v,$ we use variable \verb"edv" To represent a component
$v_i$ of the vector $v$ we use \verb"v_i".

The prefix \verb"d_" in front of a variable represents
the partial derivative $D$ and the prefix \verb"d_i" the partial derivative
$D_i.$ Thus $De$ and $D_i e$ are translated as \verb"d_e" and \verb"d_ie", respectively.

\bigskip
{\bf Header used in the proofs of the results of the paper}
\bigskip

This header was used in front of all the programs that follow.
\bigskip

{\small
\begin{verbatim}
"Header for electromagnetism"
"--"
"Define basic identities for dot product:"
ede:=1
"To avoid double entrees for dot product variables, use"
"notation that follows order of vector variables as in the list:"
"e, v, a, dot_a; thus use vda and not adv for dot product of"
"vector a with vector v. So edv is ok but not vde."
edv:=1-1/z
"--"
"Derivatives with respect to D_i"
d_it:=z*e_i
d_itau:=-z*e_i
d_iu:=-u^2*z*e_i
"--"
"Meaning of variables in the following expressions"
" un_ii=1; ee_ii=e_i*e_i"
" ev_ii=v_i*e_i=e_i*v_i"
d_ie_i:=+u*un_ii-u*z*ee_ii+u*z*ev_ii
" ea_ii=e_i*a_i"
d_iv_i:=-z*ea_ii
"--"
"Notice that edot_a_ii=e_i*a'_i"
d_ia_i:=-z*edot_a_ii
"--"
"un_i denotes here the i-th unit vector of standard base"
d_ie:=+u*un_i-u*z*e_i*e+u*z*e_i*v
d_iv:=-z*e_i*a
d_ia:=-z*e_i*dot_a
d_iz:=+u*z^2*v_i-u*z^3*e_i+u*z^2*e_i+u*z^3*vdv*e_i-z^3*eda*e_i
"D_iz:=+uz^2v_i -uz^3e_i  +uz^2e_i  +uz^3(v,v)e_i -z^3(e,a)e_i"
"--"
"Derivatives of dot products:"
d_iede:=0
d_iedv:=-u*z*edv*e_i+u*v_i+u*z*vdv*e_i-z*eda*e_i
d_ieda:=-u*z*eda*e_i+u*a_i+u*z*vda*e_i-z*eddot_a*e_i
d_ivdv:=-2*z*vda*e_i
d_ivda:=-z*ada*e_i-z*vddot_a*e_i
d_iada:=-2*z*addot_a*e_i
"--"
"Derivatives with respect to time"
d_tau:=z
d_t:=1-z
d_u:=u^2*z-u^2
d_z:=u*z-2*u*z^2+z^3*eda+u*z^3-u*z^3*vdv
"--"
"Time derivatives of vector fields"
d_v:=z*a
d_e:=-u*e+u*z*e-u*z*v
d_a:=z*dot_a
"--"
"Time derivatives for components of the vector fields"
d_v_i:=z*a_i
d_e_i:=-u*e_i+u*z*e_i-u*z*v_i
d_a_i:=z*dot_a_i
"--"
"Time derivatives of dot products:"
"Remember the ordering list: e, v, a, dot_a"
d_ede:=0
d_edv:=-u*edv+u*z*edv+u*z*vdv+z*eda
d_eda:=-u*eda+u*z*eda-u*z*vda+z*eddot_a
d_vdv:=2*z*vda
d_vda:=z*ada+z*vddot_a
d_ada:=2*z*addot_a
"End of header"
";;;;;;;;;;;;;;;;;;;;;;;;;;;;;;;;;;;;;;;;;;;;;;;;;;;;;;;;;;;;;;;;;;;;;;;;;;;"
\end{verbatim}
}
\bigskip

{\bf Check that scalar potential $\phi=uz$ satisfies the wave equation}
\bigskip

Include the header at the beginning of this file to process the program.
\bigskip

{\small
\begin{verbatim}
"Checking that scalar potential Phi=u*z satisfies the wave equation"
"--"
"Compute time derivative D(u*z)"
d_phi:=z*d_u+u*d_z
"Expanding right side of above yields"
d_phi:=-u^2*z^3*vdv+u^2*z^3-u^2*z^2+u*z^3*eda
"--"
"Compute derivative D of each term of the above sum"
tt1:=-(2*u*d_u*z^3*vdv+u^2*3*z^2*d_z*vdv+u^2*z^3*d_vdv)
tt2:=2*u*d_u*z^3+u^2*3*z^2*d_z
tt3:=-(2*u*d_u*z^2+u^2*2*z*d_z)
tt4:=d_u*z^3*eda+u*3*z^2*d_z*eda+u*z^3*d_eda
"--"
"Thus the second derivative D^2 of u*z is"
dd_phi:=tt1+tt2+tt3+tt4
"--"
"%%%%%%%%% Now compute partials D_i of u*v %%%%%%%%%"
d_iphi:=z*d_iu+u*d_iz
"Expanding right side of above yields"
d_iphi:=+u^2*z^3*vdv*e_i-u^2*z^3*e_i+u^2*z^2*v_i-u*z^3*eda*e_i
"--"
"Partial derivative D_i of each term of the above sum is"
uu1:=+2*u*d_iu*z^3*vdv*e_i+u^2*3*z^2*d_iz*vdv*e_i+u^2*z^3*d_ivdv*e_i+u^2*z^3*v~
dv*d_ie_i
uu2:=-2*u*d_iu*z^3*e_i-u^2*3*z^2*d_iz*e_i-u^2*z^3*d_ie_i
uu3:=+2*u*d_iu*z^2*v_i+u^2*2*z*d_iz*v_i+u^2*z^2*d_iv_i
uu4:=-d_iu*z^3*eda*e_i-u*3*z^2*d_iz*eda*e_i-u*z^3*d_ieda*e_i-u*z^3*eda*d_ie_i
"--"
"Thus second partial (D_i)^2 of u*z is"
dd_iphi:=uu1+uu2+uu3+uu4
"Expanding right side yields"
dd_iphi:=3*u^3*z^5*vdv^2*e_i^2-6*u^3*z^5*vdv*e_i^2+3*u^3*z^5*e_i^2-u^3*z^4*ee_~
ii*vdv+u^3*z^4*ee_ii+u^3*z^4*ev_ii*vdv-u^3*z^4*ev_ii+u^3*z^4*vdv*e_i^2+5*u^3*z~
^4*vdv*e_i*v_i-u^3*z^4*e_i^2-5*u^3*z^4*e_i*v_i+u^3*z^3*un_ii*vdv-u^3*z^3*un_ii~
+2*u^3*z^3*v_i^2-6*u^2*z^5*eda*vdv*e_i^2+6*u^2*z^5*eda*e_i^2+u^2*z^4*eda*ee_ii~
-u^2*z^4*eda*ev_ii-u^2*z^4*eda*e_i^2-5*u^2*z^4*eda*e_i*v_i-3*u^2*z^4*vda*e_i^2~
-u^2*z^3*eda*un_ii-u^2*z^3*ea_ii-u^2*z^3*e_i*a_i+3*u*z^5*eda^2*e_i^2+u*z^4*edd~
ot_a*e_i^2
"--"
"After summation over index i we get the Laplacian of Phi=u*z"
lp_phi:=+3*u^3*z^5*vdv^2*ede-6*u^3*z^5*vdv*ede+3*u^3*z^5*ede-u^3*z^4*ede*vdv+u~
^3*z^4*ede+u^3*z^4*edv*vdv-u^3*z^4*edv+u^3*z^4*vdv*ede+5*u^3*z^4*vdv*edv-u^3*z~
^4*ede-5*u^3*z^4*edv+u^3*z^3*3*vdv-u^3*z^3*3+2*u^3*z^3*vdv-6*u^2*z^5*eda*vdv*e~
de+6*u^2*z^5*eda*ede+u^2*z^4*eda*ede-u^2*z^4*eda*edv-u^2*z^4*eda*ede-5*u^2*z^4~
*eda*edv-3*u^2*z^4*vda*ede-u^2*z^3*eda*3-u^2*z^3*eda-u^2*z^3*eda+3*u*z^5*eda^2~
*ede+u*z^4*eddot_a*ede
"--"
"Check if wave equation is satisfied"
wave:=lp_phi-dd_phi
"Expanding the right side yields"
wave:=0
"Eureka! It worked."
"--"
"For the record expand the expression defining variable lp_phi"
lp_phi:=3*u^3*z^5*vdv^2-6*u^3*z^5*vdv+3*u^3*z^5+6*u^3*z^4*vdv-6*u^3*z^4-u^3*z^~
3*vdv+3*u^3*z^3-6*u^2*z^5*eda*vdv+6*u^2*z^5*eda-6*u^2*z^4*eda-3*u^2*z^4*vda+u^~
2*z^3*eda+3*u*z^5*eda^2+u*z^4*eddot_a
\end{verbatim}
}
\bigskip
{\bf Check that vector potential $A=uzv$ satisfies the wave equation}
\bigskip

{\small
\begin{verbatim}
"Check that vector potential aa=u*z*v satisfies wave equation"
"--"
"Generate Laplacian of vector potential aa=u*z*v "
"Compute first partials D_i_aa=D_i[uzv]"
"d_i_aa= d_iu*z*v   +u*d_iz*v   +u*z*d_iv"
"In expanded form the expression has 5 terms"
"d_i_aa=  -z^3*e_i*eda*u*v  +z^3*e_i*u^2*v*vdv  -z^3*e_i*u^2*v ~  "
"         -z^2*a*e_i*u      +z^2*u^2*v*v_i"
"Compute D_i derivative of each term of the above sum"
tt1:=-3*z^2*d_iz*e_i*eda*u*v-z^3*d_ie_i*eda*u*v-z^3*e_i*d_ieda*u*v-z^3*e_i*eda~
*d_iu*v-z^3*e_i*eda*u*d_iv
tt2:=+3*z^2*d_iz*e_i*u^2*v*vdv+z^3*d_ie_i*u^2*v*vdv+z^3*e_i*2*u*d_iu*v*vdv+z^3~
*e_i*u^2*d_iv*vdv+z^3*e_i*u^2*v*d_ivdv
tt3:=-3*z^2*d_iz*e_i*u^2*v-z^3*d_ie_i*u^2*v-z^3*e_i*2*u*d_iu*v-z^3*e_i*u^2*d_iv
tt4:=-2*z*d_iz*a*e_i*u-z^2*d_ia*e_i*u-z^2*a*d_ie_i*u-z^2*a*e_i*d_iu
tt5:=+2*z*d_iz*u^2*v*v_i+z^2*2*u*d_iu*v*v_i+z^2*u^2*d_iv*v_i+z^2*u^2*v*d_iv_i
"--"
d_i_d_i_aa:=tt1+tt2+tt3+tt4+tt5
"--"
"Expanded the right side of above expression yields"
d_i_d_i_aa:=3*z^5*e_i^2*eda^2*u*v-6*z^5*e_i^2*eda*u^2*v*vdv+6*z^5*e_i^2*eda*u^~
2*v+3*z^5*e_i^2*u^3*v*vdv^2-6*z^5*e_i^2*u^3*v*vdv+3*z^5*e_i^2*u^3*v+3*z^4*a*e_~
i^2*eda*u-3*z^4*a*e_i^2*u^2*vdv+3*z^4*a*e_i^2*u^2-z^4*e_i^2*eda*u^2*v+z^4*e_i^~
2*eddot_a*u*v+z^4*e_i^2*u^3*v*vdv-z^4*e_i^2*u^3*v-3*z^4*e_i^2*u^2*v*vda-5*z^4*~
e_i*eda*u^2*v*v_i+5*z^4*e_i*u^3*v*v_i*vdv-5*z^4*e_i*u^3*v*v_i+z^4*eda*ee_ii*u^~
2*v-z^4*eda*ev_ii*u^2*v-z^4*ee_ii*u^3*v*vdv+z^4*ee_ii*u^3*v+z^4*ev_ii*u^3*v*vd~
v-z^4*ev_ii*u^3*v-z^3*a*e_i^2*u^2-3*z^3*a*e_i*u^2*v_i+z^3*a*ee_ii*u^2-z^3*a*ev~
_ii*u^2-z^3*a_i*e_i*u^2*v+z^3*dot_a*e_i^2*u-z^3*ea_ii*u^2*v-z^3*eda*u^2*un_ii*~
v+z^3*u^3*un_ii*v*vdv-z^3*u^3*un_ii*v+2*z^3*u^3*v*v_i^2-z^2*a*u^2*un_ii
"--"
"To compute Laplacian(aa) we need to take sum over index i"
"remembering that sum(e_i*e_i)=ede; sum(e_i*v_i)=edv; sum(ee_ii)=ede;"
"sum(un_ii)=3; sum(v_i^2)=vdv etc."
"--"
"To this end edit previous expression to make proper changes"
"Thus Laplacian(aa) is"
lp_aa:=3*z^5*ede*eda^2*u*v-6*z^5*ede*eda*u^2*v*vdv+6*z^5*ede*eda*u^2*v+3*z^5*~
ede*u^3*v*vdv^2-6*z^5*ede*u^3*v*vdv+3*z^5*ede*u^3*v+3*z^4*a*ede*eda*u-3*z^4*a*~
ede*u^2*vdv+3*z^4*a*ede*u^2-z^4*ede*eda*u^2*v+z^4*ede*eddot_a*u*v+z^4*ede*u^3*~
v*vdv-z^4*ede*u^3*v-3*z^4*ede*u^2*v*vda-5*z^4*edv*eda*u^2*v+5*z^4*edv*u^3*v*vd~
v-5*z^4*edv*u^3*v+z^4*eda*ede*u^2*v-z^4*eda*edv*u^2*v-z^4*ede*u^3*v*vdv+z^4*ed~
e*u^3*v+z^4*edv*u^3*v*vdv-z^4*edv*u^3*v-z^3*a*ede*u^2-3*z^3*a*edv*u^2+z^3*a*ed~
e*u^2-z^3*a*edv*u^2-z^3*eda*u^2*v+z^3*dot_a*ede*u-z^3*eda*u^2*v-z^3*eda*u^2*3*~
v+z^3*u^3*3*v*vdv-z^3*u^3*3*v+2*z^3*u^3*v*vdv-z^2*a*u^2*3
"--"
"Now compute time derivative D_aa of [u*z*v]"
d_aa:=+d_u*z*v+u*d_z*v+u*z*d_v
"It expands to"
d_aa:=+u^2*z^3*v*(1-vdv)-u^2*z^2*v+eda*u*z^3*v+a*u*z^2
"--"
"Compute derivative of each term of above expression"
pp1:=+2*u*d_u*z^3*v*(1-vdv)+u^2*3*z^2*d_z*v*(1-vdv)+u^2*z^3*d_v*(1-vdv)-u^2*z^~
3*v*d_vdv
pp2:=-2*u*d_u*z^2*v-u^2*2*z*d_z*v-u^2*z^2*d_v
pp3:=+d_eda*u*z^3*v+eda*d_u*z^3*v+eda*u*3*z^2*d_z*v+eda*u*z^3*d_v
pp4:=+d_a*u*z^2+a*d_u*z^2+a*u*2*z*d_z
"--"
"The second derivative of aa with respect to time is"
dd_aa:=pp1+pp2+pp3+pp4
"--"
"It expands to"
dd_aa:=3*u^3*z^5*vdv^2*v-6*u^3*z^5*vdv*v+3*u^3*z^5*v+6*u^3*z^4*vdv*v-6*u^3*z^4~
*v-u^3*z^3*vdv*v+3*u^3*z^3*v-6*u^2*z^5*vdv*eda*v+6*u^2*z^5*eda*v-3*u^2*z^4*vdv~
*a-6*u^2*z^4*eda*v-3*u^2*z^4*vda*v+3*u^2*z^4*a+u^2*z^3*eda*v-4*u^2*z^3*a+u^2*z~
^2*a+3*u*z^5*eda^2*v+3*u*z^4*eda*a+u*z^4*v*eddot_a+u*z^3*dot_a
"--"
"D'Alembertian  of aa is"
wave_aa:=lp_aa-dd_aa
"Expanding right side of above yields:"
wave_aa:=0
"--"
"Thus the wave equation is satisfied!"
"--"
"For the record expand the right side of statement defining variable dd_aa"
dd_aa:=3*u^3*z^5*vdv^2*v-6*u^3*z^5*vdv*v+3*u^3*z^5*v+6*u^3*z^4*vdv*v-6*u^3*z^4~
*v-u^3*z^3*vdv*v+3*u^3*z^3*v-6*u^2*z^5*eda*vdv*v+6*u^2*z^5*eda*v-6*u^2*z^4*eda~
*v-3*u^2*z^4*vda*v-3*u^2*z^4*vdv*a+3*u^2*z^4*a+u^2*z^3*eda*v-4*u^2*z^3*a+u^2*z~
^2*a+3*u*z^5*eda^2*v+3*u*z^4*eda*a+u*z^4*eddot_a*v+u*z^3*dot_a
\end{verbatim}
}
\bigskip

{\bf Check that Lorentz gauge formula  $\nabla\cdot A+D\phi=0$ is satisfied.}
\bigskip

{\small
\begin{verbatim}
"Check Lorentz gauge formula Div(A)+D(Phi)=0"
"--"
"Computing D_iA_i=D_i[uzv_i]"
d_i_a_i:=d_iu*z*v_i+u*d_iz*v_i+u*z*d_iv_i
"--"
"Expanding the right side of above statement yields"
d_i_a_i:=u^2*z^3*vdv*e_i*v_i-u^2*z^3*e_i*v_i+u^2*z^2*v_i^2-u*z^3*eda*e_i*v_i-u~
*z^2*ea_ii
"--"
"Thus the divergence of the vector field A is"
div_a:=u^2*z^3*vdv*edv-u^2*z^3*edv+u^2*z^2*vdv-u*z^3*eda*edv-u*z^2*eda
"--"
"Time derivative of Phi is"
d_phi:=d_u*z+u*d_z
"--"
"Lorentz gauge is"
gauge:=div_a+d_phi
"--"
"Expanding the right side of above yields"
gauge:=0
"--"
"For the record expand the expression defining variable d_phi"
d_phi:=-u^2*z^3*vdv+u^2*z^3-u^2*z^2+u*z^3*eda
\end{verbatim}
}

\bigskip
{\bf Compare Feynman's field with Li\'enard-Wiechert's field}
\bigskip

{\small
\begin{verbatim}
"Compare Feynman's electric field ff=u^2*e+u^(-1)*D(u^2*e)+D^2e"
"with Lienard-Wiechert electric field ee=-grad(u*z)-D(u*z*v)"
"--"
"First term of the sum defining ff is"
tt1:=u^2*e
"The second term is"
tt2:=u^(-1)*(2*u*d_u*e+u^2*d_e)
"To compute the third term notice first that"
d_e:=u*z*(e-v)-e*u
"Thus for the third term we have the formula"
tt3:=d_u*z*(e-v)+u*d_z*(e-v)+u*z*(d_e-d_v)-d_e*u-e*d_u
"--"
"Thus we have for Feynman field ff"
ff:=tt1+tt2+tt3
"--"
"Now compute components ee_i of the electric field ee"
"where ee=-grad(u*z)-D(u*z*v)"
ee_i:=-(d_iu*z+u*d_iz)-(d_u*z*v_i+u*d_z*v_i+u*z*d_v_i)
"The above expands to"
ee_i:=-u^2*z^3*vdv*e_i+u^2*z^3*vdv*v_i+u^2*z^3*e_i-u^2*z^3*v_i+u*z^3*eda*e_i-u~
*z^3*eda*v_i-u*z^2*a_i
"--"
"Thus in vector notation ee is"
ee:=-u^2*z^3*vdv*e+u^2*z^3*vdv*v+u^2*z^3*e-u^2*z^3*v+u*z^3*eda*e-u*z^3*eda*v-u~
*z^2*a
difference:=ff-ee
"Expanding in the above statement the expression on the right side yields"
difference:=0
"So the electric fields coincide!"
"--"
"Formulas for magnetic fields involve cross product, programming of which is"
"cumbersome. It is easier to do it by a direct computation as in the text."

\end{verbatim}
}


\end{document}